%% file: secrecy_full_revised-4.tex
\documentclass[onecolumn, romanappendices, letterpaper]{IEEEtran}

\include{defns}

\usepackage{cite}
\usepackage{version}
\usepackage{fullpage,epsfig,graphicx,psfrag,color,subfig}
\usepackage{amsmath}
\usepackage{amssymb}
\newtheorem{lemma}{Lemma}
\newtheorem{theorem}{Theorem}
\newtheorem{proposition}{Proposition}
\newtheorem{corollary}{Corollary}

\begin{document}
\title{3-Receiver Broadcast Channels with Common and Confidential Messages}

\author{
\authorblockN{Yeow-Khiang Chia and Abbas El Gamal}\\
\authorblockA{Department of Electrical Engineering\\
Stanford University\\
Stanford, CA 94305, USA\\
Email:  ykchia@stanford.edu, abbas@ee.stanford.edu}
}

\maketitle

\begin{abstract}
This paper establishes inner bounds on the secrecy capacity regions for the general 3-receiver broadcast channel with one common and one confidential message sets. We consider two setups. The first is when the confidential message is to be sent to two receivers and kept secret from the third receiver. Achievability is established using indirect decoding, Wyner wiretap channel coding, and the new idea of generating secrecy from a publicly available superposition codebook. The inner bound is shown to be tight for a class of reversely degraded broadcast channels and when both legitimate receivers are less noisy than the third receiver. The second setup investigated in this paper is when the confidential message is to be sent to one receiver and kept secret from the other two receivers. Achievability in this case follows from Wyner wiretap channel coding and indirect decoding. This inner bound is also shown to be tight for several special cases.  
\end{abstract}

\section{Introduction}

The wiretap channel was first introduced in the seminal paper by Wyner \cite{Wyner}. He considered a 2-receiver broadcast channel where sender $X$ wishes to communicate a message to receiver $Y$ while keeping it secret from the other receiver (eavesdropper) $Z$. Wyner showed that the secrecy capacity when the channel to the eavesdropper is a degraded version of the channel to the legitimate receiver is
\[
C_{\rm s} = \max_{p(x)} (I(X;Y) -I(X;Z)).
\]
The main coding idea is to  randomly generate $2^{n(I(X:Y))}$ $x^n$ sequences  and partition them into $2^{nR}$ message bins, where $R< I(X;Y)-I(X;Z)$. To send a message, a sequence from the message bin is \emph{randomly} selected and transmitted. The legitimate receiver uniquely decodes the codeword and hence the message with high probability, while the message is kept asymptotically secret from the eavesdropper provided $R<C_{\rm S}$.

This result was extended by Csisz\'{a}r and K\"{o}rner \cite{Csiszar} to general (non-degraded) 2-receiver broadcast channels with common and confidential messages. They established the secrecy capacity region, which is the optimal tradeoff between the common and private message rates and the eavesdropper's private message equivocation rate. In the special case of no common message, their result yields the secrecy capacity for the general wiretap channel,
\[
C_{\rm s} = \max_{p(v)p(x|v)} (I(V;Y)-I(V;Z)).
\]
The achievability idea is to use Wyner's wiretap channel coding for the channel from $V$ to $Y$ by randomly selecting a $v^n$ codeword from the message bin and then sending a random sequence $X^n$ generated according to $\prod_{i=1}^np_{X|V}(x_i|v_i)$. 

The work in \cite{Csiszar} has been extended in several directions by considering different message demands and secrecy scenarios, e.g., see \cite{Ruoheng}, \cite{Khandani2008}.  However, with some notable exceptions such as \cite{Khisti} and \cite{Ulukus1}, extending the result of Csisz\'{a}r and K\"{o}rner to general discrete memoryless broadcast channels with more than two receivers has remained open, since even the capacity region without secrecy constraints for the 3-receiver broadcast channel with degraded message sets is not known in general. The secrecy setup for the 3-receiver broadcast channel also has close connections to the compound wiretap channel model (see \cite[Chapter 3]{Liang--Poor--Shamai2008b} and references therein). 
Recently, Nair and El Gamal \cite{Nair} showed that the straightforward extension of the K\"{o}rner--Marton capacity region for the 2-receiver broadcast channel with degraded message sets to more than 3 receivers is not optimal. They established an achievable rate region for the general 3-receiver broadcast channel and showed that it can be strictly larger than the straightforward extension of the K\"{o}rner--Marton region. 

In this paper, which is a much expanded version of~\cite{chia--el-gamal}, we establish inner and outer bounds on the secrecy capacity region for the 3-receivers broadcast channel with common and confidential messages. We consider two setups.
\begin{itemize}
\item \emph{2-receiver, 1-eavesdropper}: Here the confidential message is to be sent to two receivers and kept secret from the third receiver (eavesdropper). 
\item \emph{1-receiver, 2-eavesdroppers}: In this setup the confidential message is to be sent to one receiver and kept secret from the other two receivers. 
\end{itemize}

To illustrate the main coding idea in our new inner bound for the 2-receiver, 1-eavesdropper setup, consider the special case where a message $M \in [1:2^{nR}]$ is to be sent reliably to receivers $Y_1$ and $Y_2$ and kept asymptotically secret from eavesdropper $Z$. A straightforward extension of the Csisz\'{a}r--K\"{o}rner ~\cite{Csiszar} result for the 2-receiver wiretap channel yields the lower bound on the secrecy capacity 
\begin{equation}
C_{\rm S} \ge \max_{p(v)p(x|v)} \min \{I(V;Y_1)-I(V;Z), I(V;Y_2)-I(V;Z)\}. \label{eqn:1}
\end{equation}
Now, suppose $Z$ is a degraded version of $Y_1$, then from Wyner's wiretap result, we know that $(I(V;Y_1)-I(V;Z)) \le (I(X;Y_1)- I(X;Z))$ for all $p(v,x)$. However, no such inequality holds in general for the second term under the minimum. As a special case of the inner bound in Theorem~\ref{thm:1}, we show that the rate obtained by replacing $V$ by $X$ only in the first term in (\ref{eqn:1}) is achievable, that is, we establish the lower bound    
\begin{align}
C_{\rm S}\ge \max_{p(v)p(x|v)} \min \{I(X;Y_1)-I(X;Z), I(V;Y_2)-I(V;Z)\}. \label{lb1}
\end{align}
To prove achievability of (\ref{lb1}), we again randomly generate $2^{n(I(V;Y_2)-\d)}$  $v^n$ sequences  and partition them into $2^{nR}$ bins, where $R = (I(V;Y_2) - I(V;Z))$. For each $v^n$ sequence, we randomly and conditionally independently generate $2^{nI(X;Z|V)}$ $x^n$ sequences. The $v^n$ and $x^n$ sequences are revealed to all parties, including the eavesdropper. To send a message $m$, the encoder randomly chooses a $v^n$ sequence from bin $m$. It then \emph{randomly} chooses an $x^n$ sequence from the codebook for the selected $v^n$ sequence (instead of randomly generating an $X^n$ sequence as in the Csisz\'{a}r--K\"{o}rner scheme) and transmits it. Receiver $Y_2$ decodes $v^n$ directly, while receiver $Y_1$ decodes $v^n$ \emph{indirectly} through $x^n$~\cite{Nair}. In Section~\ref{sect:3}, we show through an example that this new lower bound can be strictly larger than the extended Csisz\'{a}r--K\"{o}rner lower bound. We then show in Theorem~\ref{thm:1} that this lower bound can be generalized further via Marton coding.

The rest of the paper is organized as follows. In the next section we present needed definitions. In Section~\ref{sect:3}, we provide an alternative proof of achievability for the Csisz\'{a}r--K\"{o}rner 2-receiver wiretap channel that uses superposition coding and random codeword selection instead of random generating of the transmitted codeword. This technique is used in subsequent sections to establish the new inner bounds for the 3-receiver setups. In Section~\ref{sect:4}, we present the inner bound for the 2-receiver, 1-eavesdropper case. We show that this lower bound is tight for the reversely degraded product broadcast channel and when the eavesdropper is less noisy than both legitimate receivers. In Section~\ref{sect:5}, we present inner and outer bounds for the 1-receiver, 2-eavesdropper setup for 3-receiver multilevel broadcast channel~\cite{bzt}. We show that the bounds coincide in several special cases. 

\section{Definitions and Problem Setup} \label{sect:2}
Consider a 3-receiver discrete memoryless broadcast channel with input alphabet $\mathcal{X}$, output alphabets $\mathcal{Y}_1, \mathcal{Y}_2, \mathcal{Y}_3$ and conditional pmfs $p(y_1,y_2,y_3|x)$. We investigate the following two setups.

\subsection{2-Receivers, 1-Eavesdropper} 
Here the confidential message is to be sent to receivers $Y_1$ and $Y_2$ and is to be kept secret from the eavesdropper $Y_3=Z$). A $(2^{nR_0},2^{nR_1}, n)$ code for this scenario consists of: (i) two messages $(M_0, M_1)$ uniformly distributed over $[1:2^{nR_0}]\times[1:2^{nR_1}]$; (ii) an encoder that randomly generates a codeword $X^n(m_0,m_1)$ according to the conditional pmf $p(x^n|m_0,m_1)$; and (iii) 3 decoders; the first decoder assigns to each received sequence $y_1^n$ an estimate $(\hat{M_{01}},\hat{M_{11}}) \in [1:2^{nR_0}] \times [1:2^{nR_1}]$ or an error message, the second decoder assigns to each received sequence $y_2^n$ an estimate $(\hat{M_{02}},\hat{M_{12}}) \in [1:2^{nR_0}] \times [1:2^{nR_1}]$ or an error message, and the third decoder assigns to each received sequence $z^n$ an estimate $\hat{M_{03}} \in [1:2^{nR_0}]$ or an error message. The probability of error for this scenario is defined as
\begin{align*}
P_{e1}^{(n)} &= \P\left\{\hat{M}_{0j} \neq M_0 \text{for } j=1,2,3 \text{or } \hat{M}_{1j} \neq M_1 \text{for } j=1,2 \right\}.
\end{align*}
The equivocation rate at receiver $Z$, which measures the amount of uncertainty receiver $Z$ has about message $M_1$, is defined as $H(M_1|Z^n)/n$.

A secrecy rate tuple $(R_0, R_1, R_e)$ is said to be achievable if
\begin{align*}
\lim_{n\rightarrow \infty} P_{e1}^{(n)} = 0, \text{and } \\
\liminf_{n\rightarrow \infty} \frac{1}{n}H(M_1|Z^n) \geq R_e.
\end{align*}

The \emph{secrecy capacity region} is the closure of the set of achievable rate tuples $(R_0, R_1, R_e)$.

For this setup, we also consider the special case of \emph{asymptotic perfect secrecy}, where no common message is to be sent to $Z$ and a confidential message, $M \in [1:2^{nR}]$, is to be sent to $Y_1$ and $Y_2$ only. The probability of error is as defined above with $R_0 = 0$ and $R_1 =R$. A secrecy rate $R$ is said to be achievable if there exists a sequence of $(2^{nR}, n)$ codes such that
\begin{align*}
\lim_{n\rightarrow \infty} P_{e1}^{(n)} = 0, \text{and } \\
\liminf_{n\rightarrow \infty} \frac{1}{n}H(M|Z^n)  \ge R.
\end{align*} 
The \emph{secrecy capacity}, $C_S,$ is the supremum of all achievable rates. 

\subsection{1-Receiver, 2-Eavesdroppers} 

In this setup, the confidential message is to be sent to receiver $Y_1$ and kept secret from eavesdroppers $Y_2=Z_2$ and $Y_3=Z_3$. A $(2^{nR_0},2^{nR_1}, n)$ code consists of the same message sets and encoding function as in the 2-receiver, 1-eavesdropper case. The first decoder assigns to each received sequence $y_1^n$ an estimate $(\hat{M_{01}},\hat{M_{1}}) \in [1:2^{nR_0}] \times [1:2^{nR_1}]$ or an error message, the second decoder assigns to each received sequence $z_2^n$ an estimate $\hat{M_{02}} \in [1:2^{nR_0}]$ or an error message, and the third decoder assigns to each received sequence $z_3^n$ an estimate $\hat{M_{03}} \in [1:2^{nR_0}]$ or an error message. The probability of error is 
\[
P_{e2}^{(n)} = \P\{\hat{M}_{0j} \neq M_0 \text{for } j=1,2,3 \text{or } \hat{M}_1 \neq M_1\}.
\]
The equivocation rates at the two eavesdroppers are $H(M_1|Z_2^n)/n$ and $H(M_1|Z_3^n)/n$, respectively.

A secrecy rate tuple $(R_0,R_1,R_{e2},R_{e3})$ is said to be achievable if 
\begin{align*}
\lim_{n\to \infty} P_{e2}^{(n)} &= 0, \\
\liminf_{n\to \infty} \frac{1}{n}H(M_1|Z_j^n) &\geq R_{ej},\ j=2,3. 
\end{align*}
The \emph{secrecy capacity region} is the closure of the set of achievable rate tuples $(R_0,R_1,R_{e2},R_{e3})$. For simplicity of presentation, we consider only the special class of multilevel broadcast channels~\cite{bzt}.

\section{2-receiver wiretap channel} \label{sect:3}


We first revisit the 2-receiver wiretap channel, where a confidential message is to be sent to the legitimate receiver $Y$ and kept secret from the eavesdropper $Z$. The secrecy capacity for this case is a special case of the secrecy capacity region for the broadcast channel with common and confidential messages established in~\cite{Csiszar}.

\medskip
\begin{proposition} \label{prop0}
The secrecy capacity of the 2-receiver wiretap channel is
\begin{align*}
C_{\rm S} = \max_{p(v,x)}(I(V;Y) - I(V;Z)).
\end{align*}
\end{proposition}
\medskip
In the following, we provide a new proof of achievability for this result in which the second randomization step in the original proof is replaced by a random codeword selection from a \emph{public} superposition codebook. As we will see, this proof technique allows us to use indirect decoding to establish new inner bounds for the 3-receiver wiretap channels.
\medskip

\noindent\emph{Proof of Achievability for Proposition~\ref{prop0}}:


Fix $p(v,x)$. Randomly and independently generate  sequences $v^n(l_0)$, $l_0 \in [1:2^{n\Rt}]$, each according to $\prod_{i=1}^np_V(v_i)$. Partition the set $[1:2^{n\Rt}]$ into $2^{nR}$ bins $\Bc(m)=[(m-1)2^{n(\Rt-R)}+1: m2^{n(\Rt-R)}]$, $m \in [1:2^{nR}]$. For each $l_0\in [1:2^{n\Rt}]$, randomly and conditionally independently generate sequences $x^n(l_0, l_1)$, $l_1 \in [1:2^{n\Rt_1}]$, each according to $\prod_{i=1}^np_{X|V}(x_i|v_i)$. The  codebook $\{(v^n(l_0),x^n(l_0,l_1))\}$ is revealed to all parties. To send the message $m$, an index $L_0 \in \Bc(m)$ is selected uniformly at random (as in Wyner's original proof). The encoder then randomly and independently selects an index $L_1$ and transmits $x^n(L_0,L_1)$. Receiver $Y$ decodes $L_0$ by finding the unique index $\lh_0$ such that $(v^n(\lh_0),y^n) \in \aep$. By the law of large numbers and the packing lemma \cite[Chapter 3]{El-Gamal--Kim2010}, the average probability of error approaches zero as $n \to \infty$ if $\Rt < (V;Y) -\d(\e)$.

We now show that $I(M;Z^n|\Cc) \le n\d(\e)$. Considering the mutual information between $Z^n$ and $M$, averaged over the random codebook $\Cc$, we have
\begin{align}
I(M;Z^n|\Cc) &= I(M, L_0, L_1; Z^n|\Cc) - I(L_0,L_1;Z^n|M,\Cc)\nonumber\\
&\stackrel{(a)}{\leq} I(X^n;Z^n|\Cc) - H(L_0,L_1|M,\Cc) + H(L_0,L_1|M,Z^n,\Cc) \nonumber\\
& \leq \sum_{i=1}^n I(X_i;Z_i|\Cc) - n(\Rt-R)-n\Rt_1 + H(L_0,L_1|M,Z^n,\Cc) \nonumber\\
& \leq nI(X;Z) - n(\Rt+\Rt_1-R) + H(L_0|M,Z^n,\Cc) + H(L_1|L_0, Z^n,\Cc). \label{eqn:0}
\end{align}
$(a)$ follows since $(M,L_0, L_1, \Cc) \to X^n \to Z^n$ from the discrete memoryless property of the channel. The last step follows from  follows since $H(Z_i|\Cc) \leq H(Z_i) = H(Z)$ and $H(Z_i|X_i, \Cc) = \sum_{\Cc} p(\cc)p(x_i|\cc)H(Z|x_i,\cc) = \sum_{\Cc}p(\cc)p(v_i|\cc)H(Z|x_i) = H(Z|X)$ It remains to upper bound $H(L_0|M,Z^n,\Cc)$ and $H(L_1|L_0, Z^n,\Cc)$.  By symmetry of codebook construction, we have
\begin{align*}
H(L_0|M,Z^n, \Cc) &= 2^{-nR}\sum_{m=1}^{2^{nR}}H(L_0|M=m,Z^n,\Cc) \nonumber \\
&= H(L_0|Z^n, M=1, \Cc),\\
H(L_1|L_0, Z^n, \Cc) & = 2^{-n\Rt}\sum_{l_0}H(L_1|L_0 = l_0,Z^n, \Cc) \nonumber \\
& = H(L_1|L_0 = 1,v^n(1), Z^n, \Cc). 
\end{align*}

To further bound these terms, we use the following key lemma. 
\medskip

\begin{lemma} \label{lem1}
Let $(U,V,Z) \sim p(u,v,z)$, $S \ge 0$ and $\e >0$. Let $U^n$ be a random sequence distributed according to $\prod_{i=1}^np_U(u_i)$. Let $V^n(l)$, $l \in [1:2^{nS}]$, be a set of random sequences that are conditionally independent given $U^n$ and each distributed according to $\prod_{i=1}^n p_{V|U}(v_i|u_i)$, and let $\Cc= \{U^n, V^n(l)\}$. Let $L \in [1:2^{nS}]$ be a random index with an arbitrary probability mass. Then, if $\P\{(U^n,V^n(L), Z^n)\in \aep\}\to 1$ as $n\to \infty$ and $S \geq I(V;Z|U) + \d(\e)$, there exists a $\d'(\e)>0$, where $\d'(\e) \to 0$ as $\e \to 0$, such that, for $n$ sufficiently large,
\begin{align*}
H(L|Z^n,U^n, \Cc) \leq n(S - I(V;Z|U)) +n\d'(\e).
\end{align*}
\end{lemma} 
\medskip

\begin{figure}[h]
\begin{center}
\psfrag{u}[c]{\large $U^n$}
\psfrag{z}[c]{\large $Z^n$}
\psfrag{v1}[c]{$V^n(1)$}
\psfrag{v2}[c]{$V^n(2)$}
\psfrag{vr}[c]{$V^n(2^{nS})$}
\psfrag{vl}[c]{$V^n(L)$}
\scalebox{0.7}{\includegraphics{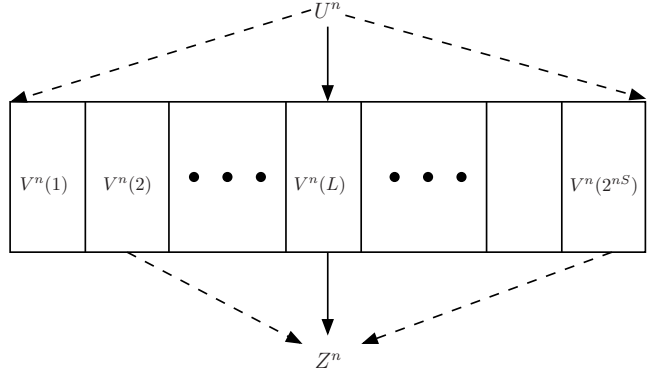}}
\caption{Structure of random sequences in Lemma \ref{lem1}. $V^n(l)$ is generated according to $\prod_{i=1}^n p_{V|U}(v_i|u_i)$. Solid arrows represent the sequence pair $(U^n, V^n(L), Z^n)$ while the dotted arrows to $Z^n$ represent the other $V^n$ sequences jointly typical with the $(U^n, Z^n)$ pair. Lemma \ref{lem1} gives an upper bound on the number of $V^n$ sequences that can be jointly typical with a $(U^n, Z^n)$ pair.} \label{fig:lem}
\end{center}
\end{figure}

The proof of this lemma is given in Appendix~\ref{appen1}. An illustration of the random sequence structure is given in Figure \ref{fig:lem}.

Now, returning to~\eqref{eqn:0}, we note that $\P\{(V^n(L_0),X^n(L_0,L_1),Z^n)\in \aep\}\to 1$ as $n\to \infty$ by law of large numbers. Hence, we can apply Lemma~\ref{lem1} to obtain
\begin{align}
H(L_0|Z^n,M=1, \Cc) & \le n((\Rt-R)-I(V;Z)) + n\d(\e), \label{eq:1}\\
H(L_1|L_0 = 1, V^n,Z^n, \Cc) &\le n(\Rt_1-I(X;Z|V)) + n\d(\e), \label{eq:2}
\end{align}
if $\Rt - R \ge I(V;Z) + \d(\e)$ and $\Rt_1 \ge I(X;Z|V) + \d(\e)$. Substituting from inequalities \eqref{eq:1} and \eqref{eq:2} into \eqref{eqn:0} shows that $I(M;Z^n|\Cc)\le 2n\d(\e)$. We then recover the original asymptotic secrecy rate by noting that the constraint of $\Rt_1 \ge I(X;Z|V)$ is not tight. This completes the proof of Proposition \ref{prop0}.
\medskip

\noindent \emph{Remark 3.1}: In the proof of Proposition~\ref{prop0} in~\cite{Csiszar}, the encoder transmits a randomly generated codeword $X^n \sim \prod_{i=1}^n p_{X|V}(x_i|v_i)$. Although replacing random $X^n$ generation by superposition coding and random codeword selection in our alternative proof does not increase the achievable secrecy rate for the 2-receiver wiretap channel, it can increase the rate when there are more than one legitimate receiver, as we show in the next sections. 

\section{2-receivers, 1-eavesdropper wiretap channel} \label{sect:4}

We establish an inner bound on the secrecy capacity for the 3-receiver wiretap channel with one common and one confidential message when the confidential message is to be sent to receivers $Y_1$ and $Y_2$ and kept secret from receiver $Z$. 
In the following subsection, we  consider the case where $M_0=\emptyset$ and $M_1=M\in [1:2^{nR}]$ is to be kept asymptotically secret from $Z$. This result is then extended in Subsection~\ref{subsect1}  to establish an inner bound on the secrecy capacity region.

\subsection{Asymptotic perfect secrecy}
We establish the following lower bound on secrecy capacity for the case where a confidential message is to be sent to receivers $Y_1$ and $Y_2$ and kept secret from the eavesdropper $Z$.
\medskip

\begin{theorem} \label{thm:1}
The secrecy capacity of the 2-receiver, 1-eavesdropper setup with one confidential message and asymptotic secrecy is lower bounded as follows
\begin{align*}
C_{\rm S} &\ge \min \{I(V_0, V_1; Y_1|Q) - I(V_0, V_1;Z|Q), I(V_0, V_2; Y_2|Q) - I(V_0, V_2;Z|Q)\} 
\end{align*}
for some $p(q,v_0,v_1,v_2,x)=p(q, v_0)p(v_1,v_2|v_0)p(x|v_1,v_2, v_0)$ such that $I(V_1,V_2;Z|V_0) \leq I(V_1;Z|V_0)+ I(V_2;Z|V_0)-I(V_1;V_2|V_0)$.
\end{theorem}
In addition to superposition coding and the new coding idea discussed in the previous section, Theorem \ref{thm:1} also uses Marton coding \cite{Marton}. 

\medskip
For clarity, we first establish the following Corollary \ref{coro1}.
\medskip
\begin{corollary} \label{coro1}
The secrecy capacity for the 2-receiver, 1-eavesdropper with one confidential message and asymptotic secrecy is lower bounded as follows
\begin{align*}
C_{\rm S} \geq \max_{p(q)p(v|q)p(x|v)}\min\{I(X;Y_1|Q)-I(X;Z|Q), I(V;Y_2|Q)-I(V;Z|Q)\}.
\end{align*}
\end{corollary}

\medskip

\noindent \emph{Remark 4.1}: Consider the case where $X \to Y_1 \to Z$ form a Markov chain. Then, we can show that Theorem~\ref{thm:1} reduces to Corollary~\ref{coro1}, i.e., the achievable secrecy rate is not increased by using Marton coding when $X \to Y_1 \to Z$ (or $X \to Y_2 \to Z$ by symmetry) form a Markov chain. To see this, note that $(I(X;Y_1|Q)-I(X;Z|Q)) \geq (I(V_1, V_0;Y_1|Q)-I(V_1,V_0;Z|Q))$ for all $V_1$ if $X \to Y_1 \to Z$. Next, note that we can set $V= (V_0,V_2)$ in Corollary \ref{coro1} to obtain the rate in Theorem \ref{thm:1}.

\medskip

\noindent\emph{Proof of Corollary~\ref{coro1}}:

\noindent \emph{Codebook generation}: Randomly and independently generate the time-sharing sequence $q^n$ according to $\prod_{i=1}^np_Q(q_i)$. Next, randomly and conditionally independently generate $2^{n\Rt}$ sequences $v^n(l_0)$, $l_0 \in [1:2^{n\Rt}]$, each according to $\prod_{i=1}^n p_{V|Q}(v_i|q_i)$. Partition the set $[1:2^{n\Rt}]$ into $2^{nR}$ equal size bins $\Bc(m)$, $m \in [1:2^{nR}]$. For each $l_0$, conditionally independently generate sequences $x^n(l_0,l_1)$, $l_1 \in [1:2^{n\Rt_1}]$, each according to $\prod_{i=1}^n p_{X|V}(x_i|v_i)$. 

\noindent \emph{Encoding}: To send a message $m\in [1:2^{nR}]$, randomly and independently choose an index $L_0\in \Cc(m)$ and an index $L_1 \in [1:2^{n\Rt_1}]$, and send $x^n(L_0,L_1)$.

\noindent \emph{Decoding}: Assume without loss of generality that $L_0 =1$ and $m=1$. Receiver $Y_2$ finds $L_0$, and hence $m$, via joint typicality decoding. By the law of large number and the packing lemma, the probability of error approaches zero as $n \to \infty$ if
\begin{align*}
\Rt &< I(V;Y_2|Q)- \d(\e).
\end{align*}

Receiver $Y_1$ finds $L_0$ (and hence $m$) via indirect decoding. That is, it declares that $\hat{L}_0$ is sent if it is the unique index such that $(q^n,v^n(\hat{L}_0), x^n(\hat{L}_0, l_1),Y_1^n)\in \aep$ for some $l_1 \in [1:2^{n\Rt_1}]$. To analyze the average probability of error $\P(\Ec)$, define the error events
\begin{align*}
\Ec_{10} &= \{(Q^n, X^n(1, 1), Y_1^n) \notin \aep\},\\
\Ec_{11} &= \{(Q^n, X^n(l_0, l_1),Y_1^n) \in \aep \text{for some } l_0 \neq 1\}.
\end{align*}
Then, by union of events bound the probability of error is upper bounded as
\[
\P(\Ec) \leq \P\{\Ec_{10}\} + \P\{\Ec_{11}\}.
\]

Now by law of large numbers, $\P\{\Ec_{10}\}\to 0$ as $n\to \infty$. Next consider
\begin{align*}
\P\{\Ec_{11}\} &\le \sum_{l_0\neq 1}\sum_{l_1} \P\{(Q^n, V^n(l_0), X^n(l_0,l_1), Y_1^n) \in \aep \} \\
& \le \sum_{l_0\neq 1}\sum_{l_1} 2^{-n(I(V,X;Y_1|Q)-\d(\e))} \\
& \le 2^{n(\Rt + \Rt_1 - I(V,X;Y_1|Q)+\d(\e))}.
\end{align*}
Hence, $\P\{\Ec_{11}\} \to 0$ as $n \to \infty$ if
\begin{align*}
\Rt+\Rt_1 < I(X;Y_1|Q)-\d(\e).
\end{align*}

\noindent \emph{Analysis of equivocation rate}: To bound the equivocation rate term $H(M|Z^n, \Cc)$, we proceed as before and show that the $I(M;Z^n|C) \le 2n\d(\e)$. Note that the only difference between this case and the analysis for the 2-receiver case in Section~\ref{sect:2} is the addition of the time-sharing random variable $Q$. Since \\ $\P\{(Q^n, V^n(L_0),X^n(L_0,L_1),Z^n)\in \aep\} \to 1$ as $n \to \infty$, we can apply Lemma~\ref{lem1} (with the addition of the time sharing random variable). Following the analysis in Section~\ref{sect:2}, it is easy to see that $I(M;Z^n|C) \le 2n\d(\e)$ if
\begin{align*}
\Rt - R &\ge I(V;Z|Q)+ \d(\e), \\
\Rt_1 &\ge I(X;Z|V)+\d(\e).
\end{align*}
Finally, using Fourier--Motzkin elimination on the set of inequalities completes the proof of achievability.

\medskip

Before proving Theorem~\ref{thm:1}, we show through an example that the lower bound in Corollary~\ref{coro1} can be strictly larger than the rate of the straightforward extension of the Csisz\'{a}r--K\"{o}rner scheme to the 2-receiver, 1-eavesdropper setting,
\begin{align}
R_{CK} = \max_{p(q)p(v|q)p(x|v)}\min \{I(V;Y_1|Q) - I(V;Z|Q), I(V;Y_2|Q) - I(V;Z|Q)\}. \label{eqn:2}.
\end{align} 
Note that Theorem \ref{thm:1} includes $R_{CK}$ as a special case (through setting $V_0 = V_1= V_2 = V$ in Theorem \ref{thm:1}).

\medskip

\begin{figure}[ht] 
\centering
\psfrag{p1}{$1/2$}
\psfrag{p2}{$1/3$}
\psfrag{p3}{$1/2$}
\psfrag{p4}{$2/3$}
\psfrag{p5}{$1/2$}
\psfrag{p6}{$1/2$}
\psfrag{0}{$0$}
\psfrag{1}{$1$}
\psfrag{E}{$E$}
\psfrag{y21}{$Y_{21}$}
\psfrag{y11}{$Y_{11}$}
\psfrag{y12}{$Y_{12}$}
\psfrag{y32}{$Z_{2}$}
\psfrag{y31}{$Z_{1}$}
\psfrag{x1}{$X_1$}
\psfrag{x2}{$X_2$}
\scalebox{0.75}{\includegraphics{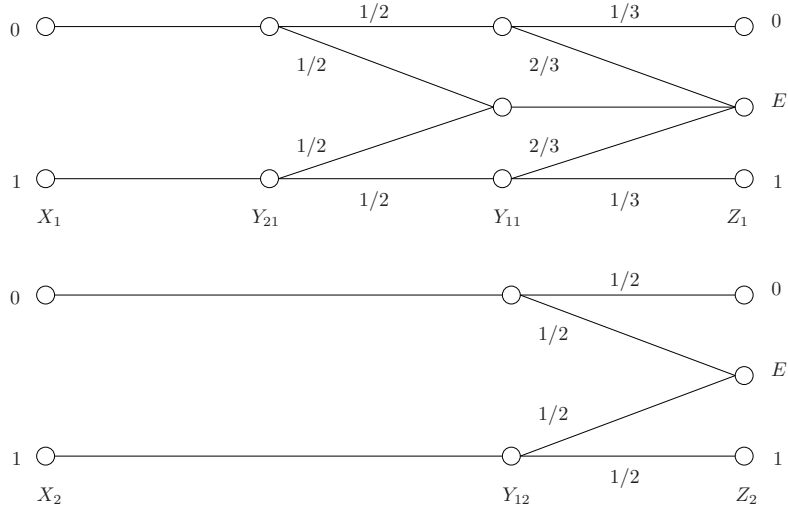}} 
\caption{Multilevel broadcast channel}\label{fig1}
\end{figure}

\noindent\emph{Example}:  Consider the multilevel product broadcast channel example~\cite{Nair} in Figure~\ref{fig1}, where $\mathcal{X}_1 = \mathcal{X}_2=\mathcal{Y}_{12} = \mathcal{Y}_{21} = \{0,1\}$, and $\mathcal{Y}_{11} = \mathcal{Z}_{1} = \mathcal{Z}_{2} = \{0,E,1\}$. The channel conditional probabilities are  specified in Figure~\ref{fig1}.

In Appendix \ref{appen2}, we show that $R_{CK} < 5/6$. In contrast, using Corollary \ref{coro1}, we can achieve a rate of $5/6$, which shows that the rate given in Theorem \ref{thm:1} can be strictly larger than using the straightforward extension of the Csisz\'{a}r--K\"{o}rner scheme.

We now turn to the proof of Theorem~\ref{thm:1}, which utilizes Marton coding in addition to the ideas already introduced.

\noindent\emph{Proof of Theorem~\ref{thm:1}}:

\noindent\emph{Codebook generation}: 
Randomly and independently generate a time-sharing sequence $q^n$ according to $\prod_{i=1}^np_Q(q_i)$. Randomly and conditionally independently generate sequences $v_0^n(l_0)$, $l_0 \in [1:2^{n\Rt}]$, each according to $\prod_{i=1}^np_{V_0|Q}(v_{0i}|q_i)$. Partition the set $[1:2^{n\Rt}]$ into $2^{nR}$ bins, $\Bc(m)$, $m \in [1:2^{nR}]$ as before. For each $l_0$, randomly and conditionally independently generate sequences $v_1^n(l_0,t_1)$, $t_1 \in [1:2^{nT_1}]$, each according to $\prod_{i=1}^n p_{V_1|V_0}(v_{1i}|v_{0i})$. Partition the set $[1:2^{nT_1}]$ into $2^{n\Rt_1}$ equal size bins, $\Bc(l_0,l_1)$. Similarly, for each $l_0$, generate  sequences $v_2^n(l_0, t_2)$,  $t_2 \in [1:2^{nT_2}]$, each according to $\prod_{i=1}^np_{V_2|V_0}(v_{2i}|v_{0i})$, and partition $[1:2^{nT_1}]$  into $2^{n\Rt_2}$ equal size bins, $\Bc(l_0,l_2)$. Finally, for each product bin $\Bc(l_0,l_1) \times \Bc(l_0,l_2)$, find a jointly typical sequence pair $(v_1^n(l_0, t_1(l_0, l_1)), v_2^n(l_0, t_2(l_0,l_2))$. If there is more than one such pair, randomly and uniformly pick one of them. This encoding step succeeds with probability of error that approaches zero as $n \to \infty$, if~\cite{Abbas2}
\begin{align*}
\Rt_1 + \Rt_2 < T_1 + T_2 - I(V_1;V_2|V_0)-\d(\e).
\end{align*}

\noindent\emph{Encoding}:
To send message $m$, the encoder first randomly chooses an index $L_0\in \Bc(m)$. It then randomly chooses a product bin indices $(L_1,L_2)$ and  selects the jointly typical sequence pair \\ $(v_1^n(L_0, t_1(L_0, L_1)), v_2^n(L_0, t_2(L_0,L_2))$. Finally, the encoder generates a codeword $X^n$ at random according to $\prod_{i=1}^np_{X|V_0,V_1,V_2}(x_i|v_{0i}, v_{1i},v_{2i})$ and transmits it. 
\medskip

\noindent\emph{Decoding and analysis of the probability of error}:
Receiver $Y_1$ decodes $L_0$ and hence $m$ indirectly by finding the unique index $\lh_0$ such that $(v^n_0(\lh_0),v_1^n(\lh_0,t_1),y_1^n) \in \aep$ for some $t_1\in [1:2^{nT_1}]$. Similarly, receiver $Y_2$ finds $L_0$ (and hence $m$) indirectly by finding the unique index $\lh_0$ such that $(v^n_0(\lh_0),v^n_2(\lh_0, T_2)) \in \aep$ for some $l_2 \in [1:2^{nT_2}]$. Following the analysis given earlier, it is easy to see that these steps succeed with probability of error that approaches zero as $n \to \infty$ if
\begin{align*}
\Rt + T_1 &< I(V_0,V_1;Y_1|Q)-\d(\e),\\
\Rt + T_2 &< I(V_0,V_1;Y_2|Q)-\d(\e).
\end{align*}
\noindent\emph{Analysis of equivocation rate}: A codebook $\cc$ induces a joint pmf on $(M,L_0, L_1, L_2, V_0^n, V_1^n, V_2^n,Z^n)$ of the form\\ $p(m,l_0, l_1, l_2, v_0^n,v_1^n,v_2^n z^n|c) = 2^{-n(\Rt + \Rt_1 + \Rt_2)}p(v_0^n, v_1^n,v_2^n|l_0,l_1,l_2,c)\prod_{i=1}^np_{Z|V_0, V_1,V_2}(z_i|v_{0i},v_{1i},v_{2i})$. We again analyze the mutual information between $M$ and $(Z^n,Q^n)$, averaged over codebooks.{\allowdisplaybreaks
\begin{align}
I(M;Z^n,Q^n|\Cc) &= I(M;Z^n|Q^n, \Cc) \nonumber \\
& = I(T_1(L_0,L_1), T_2(L_0,L_1), L_0, M; Z^n| Q^n, \Cc)\nonumber \\
&\qquad  - I(T_1(L_0,L_1), T_2(L_0,L_2), L_0; Z^n|M, Q^n, \Cc) \nonumber \\
& \leq I(V_0^n, V_1^n, V_2^n; Z^n| Q^n, \Cc) - I(L_0; Z^n |M, Q^n, \Cc) \nonumber \\
& \qquad  - I(T_1(L_0,L_1), T_2(L_0,L_2); Z^n|L_0,  Q^n, \Cc) \nonumber\\
& \leq nI(V_0,V_1, V_2; Z|Q) -  H(L_0 |M, Q^n, \Cc) + H(L_0 |M, Q^n, Z^n, \Cc) \nonumber\\
&\qquad \qquad - I(T_1(L_0,L_1), T_2(L_0,L_2); Z^n|L_0,Q^n, \Cc) + n\d(\e). \label{T1e1}
\end{align}
In the last step, we bound the term $I(V_0^n, V_1^n, V_2^n; Z^n|Q^n, \Cc)$ by the following argument, which is an extension of a similar argument in \cite{Wyner}. For simplicity of notation, let $V = (V_0, V_1, V_2)$. We wish to show that $I(V^n;Z^n|Q^n, \Cc) \le nI(V;Z) + n\d(\e)$. Note that $X$ is generated according to $p(x_i|v_i)$. Define $E=1$ if $(q^n, v^n, z^n)$ are not jointly typical and $0$ otherwise, and $N(v)= |\{V_i: V_i = v\}|$. Then,
\begin{align*}
I(V^n;Z^n|\Cc, Q^n) &\le 1 + \P(E=0) I(V^;Z^n|\Cc, E = 0, Q^n) +  \P(E=1) I(V^;Z^n|\Cc, E = 1,Q^n) \\
& \le 1 + \P(E=0) I(V^;Z^n|\Cc, E = 0, Q^n)  \\
& \quad +  \P(E=1) n\log |\Zc| -\P(E=1)H(Z^n|\Cc,V^n,Q^n, E=1) \\
& = 1 + \P(E=0) (H(Z^n|\Cc, E = 0,Q^n) - H(Z^n|\Cc,V^n,Q^n) +  \P(E=1) n\log |\Zc|.
\end{align*}
Note that $H(Z^n|\Cc, E = 0,Q^n)  \le nH(Z|Q) + n\d(\e)$. For $H(Z^n|\Cc,V^n,Q^n,E=0) =H(Z^n|\Cc,V^n,E=0)$, we have
\begin{align*}
H(Z^n|\Cc, E = 0, V^n) & \ge \sum_{c, v^n \in \aep} \P(V^n = v^n, \Cc = c)H(Z^n|\Cc = c,V^n = v^n) \\
& =  \sum_{c, v^n \in \aep} \P(V^n = v^n, \Cc = c)\sum_{i=1}^nH(Z_i|\Cc = c,  V^n = v^n, Z^{i-1}) \\
& \stackrel{(a)}{=} \sum_{c, v^n \in \aep} \P(V^n = v^n, \Cc = c)\sum_{i=1}^nH(Z_i|V_i = v_i) \\
& = \sum_{c, v^n \in \aep} \P(V^n = v^n, \Cc = c)\sum_{v \in \mathcal{V}}N(v)H(Z|V = v) \\
&\stackrel{(b)}{\ge} \sum_{c, v^n \in \aep} \P(V^n = v^n, \Cc = c)\left(\sum_{v \in \mathcal{V}}n(p(v)- \d(\e))H(Z|V = v) \right) \\
& \ge nH(Z|V) -n \d'(\e),
\end{align*}
where $(a)$ follows since given $V_i$, $X_i$ is generated randomly according to $p(x_i|v_i)$ and since the channel is memoryless, $Z_i$ is independent of all other random variables, and $(b)$ follows since $v^n$ is typical, which implies that $N(v) \ge np(v) - n\d(\e)$. Finally, since the coding scheme satisfies the encoding constraints, the proof is completed by noting that $\P(E = 1) \to 0$ as $n \to \infty$ by the law of large numbers and the mutual covering lemma in~\cite[Chapter 9]{El-Gamal--Kim2010}).

We now bound each remaining terms in inequality (\ref{T1e1}) separately. Note that  
\begin{align}
H(L_0 |M, Q^n, \Cc) &= n(\Rt-R), \label{T1e2}\\
H(L_0 |M, Q^n, Z^n, \Cc) &\stackrel{(a)}{\leq} n(S_0-R -I(V_0;Z|Q)+ \d(\e)), \label{T1e3}
\end{align}
where $(a)$ follows by similar steps to the proof of Corollary~\ref{coro1} and application of Lemma~\ref{lem1}, which holds if $\P\{(Q^n, V_0^n(L_0), Z^n)\in \aep\} \to 1$ as $n\to \infty$ and $S_0 - R \geq I(V_0;Z|Q)$. The first condition follows since \\ $\P\{(Q^n, V_0^n(L_0), V_1^n(L_0,T_1(L_0,L_1)), V_2^n(L_0,T_2(L_0,L_2)), Z^n)\in \aep\} \to 1$ as $n\to \infty$.
Next, consider 
\begin{align}
& I(T_1(L_0,L_1), T_2(L_0,L_2); Z^n|L_0,Q^n, \Cc) \nonumber\\
&\qquad  = H(T_1(L_0,L_1), T_2(L_0,L_2)|L_0,Q^n, \Cc) - H(T_1(L_0,L_1), T_2(L_0,L_2)|L_0,Q^n, Z^n, \Cc) \nonumber\\
&\qquad  \stackrel{(a)}{=} H(L_1, L_2|L_0,Q^n, \Cc) - H(T_1(L_0,L_1), T_2(L_0,L_2)|L_0,Q^n, Z^n, \Cc) \nonumber \\
&\qquad \ge H(L_1, L_2|L_0,Q^n, \Cc) - H(T_1(L_0,L_1)|L_0,Q^n, Z^n, \Cc) - H(T_2(L_0,L_2)|L_0,Q^n, Z^n, \Cc), \label{T1e4} 
\end{align}
where $(a)$ holds since given the codebook $\cc$ and $L_0$, $(T_1, T_2)$ is a one-to-one function of $(L_1,L_2)$. Now,
\begin{align}
H(L_1, L_2|L_0,Q^n, \Cc) &= n(\Rt_1+\Rt_2), \label{T1e5}\\
H(T_1(L_0,L_1)|L_0,Q^n, Z^n, \Cc) & \stackrel{(b)}{\leq} n(T_1 - I(V_1;Z|V_0) + \d(\e)), \label{T1e6}\\
H(T_2(L_0,L_2)|L_0,Q^n, Z^n, \Cc) & \stackrel{(c)}{\leq} n(T_2 - I(V_2;Z|V_0) + \d(\e)), \label{T1e7}
\end{align}
where $(b)$ and $(c)$ come from the following analysis. First consider
\begin{align*}
H(T_1(L_0,L_1)|L_0,Q^n, Z^n, \Cc) &= H(T_1(L_0,L_1)|v_0^n(L_0),Q^n, Z^n, L_0,\Cc) \\
& \le H(T_1(L_0,L_1)|V_0^n, Z^n). 
\end{align*}
We now upper bound the term $H(T_1(L_0,L_1)|V_0^n, Z^n)$. 

Since $\P\{(Q^n, V_0^n(L_0), V_1^n(L_0,T_1(L_0,L_1)), V_2^n(L_0,T_2(L_0,L_2)), Z^n)\in \aep\} \to 1$ as $n\to \infty$, \\$\P\{(V_0^n(L_0), V_1^n(L_0,T_1(L_0,L_1)),Z^n)\in \aep\} \to 1$ as $n \to \infty$. We can therefore apply Lemma 1 to obtain
\begin{align*}
H(T_1(L_0,L_1)|L_0,Q^n, Z^n, \Cc) & \le n(T_1 - I(V_1;Z|V_0) + \d(\e)), 
\end{align*}
if $T_1 \ge I(V_1;Z|V_0) + \d(\e)$.

The term $H(T_2(L_0,L_2)(L_0,L_2)|L_0,Q^n, Z^n, \Cc)$ can be bound using the same steps to give
\begin{align*}
H(T_2(L_0,L_2)|L_0,Q^n, Z^n, \Cc) & \le n(T_2 - I(V_2;Z|V_0) + \d(\e)), 
\end{align*}
if $T_2 \ge I(V_2;Z|V_0) + \d(\e)$.}

Substituting from (\ref{T1e5}), (\ref{T1e6}), and (\ref{T1e7}) into (\ref{T1e4}) yields
\begin{align}
& I(T_1(L_0,L_1), T_2(L_0,L_2); Z^n|L_0,Q^n, \Cc) \nonumber \\
& \qquad \ge n(\Rt_1+\Rt_2) -n(T_1 - I(V_1;Z|V_0) + \d(\e)) -n(T_2 - I(V_2;Z|V_0) + \d(\e)). \label{eqnx}
\end{align}
Substituting inequality (\ref{eqnx}), together with (\ref{T1e2}) and (\ref{T1e3}) into (\ref{T1e1}) then yields
\begin{align*}
I(M;Z^n|Q^n,\Cc) &\le n(I(V_1;V_2;Z|V_0) + T_1 + T_2 -\Rt_1 - \Rt_2 - I(V_1;Z|V_0) - I(V_2;Z|V_0) + 3\d(\e)).
\end{align*}
Hence, $I(M;Z^n|Q^n,\Cc) \le 3n\d(\e)$ if 
\begin{align*}
I(V_1;V_2;Z|V_0) + T_1 + T_2 -\Rt_1 - \Rt_2 - I(V_1;Z|V_0) - I(V_2;Z|V_0) \le 0.
\end{align*}
In summary, the rate constraints arising from the analysis of equivocation rate are 
\begin{align*}
S_0 -R &\ge I(V_0;Z|Q), \\
T_1 &\ge I(V_1;Z|V_0), \\
T_2 &\ge I(V_2;Z|V_0), \\
T_1 + T_2 -\Rt_1 - \Rt_2  &\le I(V_1;Z|V_0) + I(V_2;Z|V_0) - I(V_1;V_2;Z|V_0).
\end{align*}
Applying Fourier-Motzkin elimination gives
\begin{align*}
R &<  I(V_0, V_1; Y_1|Q) - I(V_0, V_1;Z|Q), \\
R &< I(V_0, V_2; Y_2|Q) - I(V_0, V_2;Z|Q), \\
2R &<  I(V_0, V_1; Y_1|Q) + I(V_0, V_2; Y_2|Q) - 2I(V_0;Z|Q) -I(V_1;V_2|V_0) 
\end{align*}
for some $p(q,v_0,v_1,v_2,x)=p(q, v_0)p(v_1,v_2|v_0)p(x|v_1,v_2, v_0)$ such that $I(V_1,V_2;Z|V_0) \leq I(V_1;Z|V_0)+ I(V_2;Z|V_0)-I(V_1;V_2|V_0)$. 

The proof of Theorem~\ref{thm:1} is then completed by observing that the third inequality is redundant. This is seen by summing the first two inequalities to yield
\begin{align*}
2R &\le I(V_0, V_1; Y_1|Q) - I(V_0, V_1;Z|Q) + I(V_0, V_2; Y_2|Q) - I(V_0, V_2;Z|Q) \\
& = I(V_0, V_1; Y_1|Q) - I(V_0, V_1;Z|Q) + I(V_0, V_2; Y_2|Q) - 2I(V_0;Z|Q) - I(V_1;Z|V_0) - I(V_2;Z|V_0).
\end{align*}
This inequality is at least as tight as the third inequality because of the constraint on the pmf. This completes the proof of Theorem~\ref{thm:1}.

\subsection*{Special Cases:} 

We consider several special cases in which the inner bound in Theorem~\ref{thm:1} is tight.
\medskip

\noindent \emph{Reversely Degraded Product Broadcast Channel}: As an example of Theorem~\ref{thm:1}, consider the reversely degraded product broadcast channel with sender $X = (X_1, X_2 \ldots, X_k)$, receivers $Y_j = (Y_{j1}, Y_{j2} \ldots, Y_{jk})$ for $j =1, 2,3$, and conditional probability mass functions $p(y_1,y_2,z|x) = \prod_{l=1}^k p(y_{1l}, y_{2l}, z_{l}|x_l)$. In \cite{Khisti}, the following lower bound on secrecy capacity is established 
\begin{align}
C_{\rm S} &\ge \min_{j\in \{1,2\} } \sum_{l=1}^k [I(U_l; Y_{jl}) - I(U_l; Z_{l})]^+. \label{eqn2}
\end{align}
for some $p(u_1,\ldots,u_k,x) = \prod_{l=1}^k p(u_l)p(x_l|u_l)$. Furthermore, this lower bound is shown to be optimal when the channel is reversely degraded (with $U_l = X_l$), i.e., when each sub-channel is degraded but not necessarily in the same order. We can show that this result is a special case of Theorem~\ref{thm:1}. Define the sets of $l$ indexes: $\mathcal{C}= \{l :\ I(U_l; Y_{1l}) - I(U_l; Z_{l}) \geq 0, I(U_l; Y_{2l}) - I(U_l; Z_{l}) \geq 0\}$, $\mathcal{A}= \{l :\ I(U_l; Y_{1l}) - I(U_l; Z_{l}) \geq 0 \}$ and $\mathcal{B}= \{l :\ I(U_l; Y_{2l}) - I(U_l; Z_{l}) \geq 0 \}$. Now, setting $V_0 = \{U_l: l \in \mathcal{C}\}$, $V_1= \{U_l: l \in \mathcal{A}\}$, and $V_2= \{U_l: l \in \mathcal{B}\}$ in the rate expression of Theorem~\ref{thm:1} yields (\ref{eqn2}). Note that the constraint in Theorem~\ref{thm:1} is satisfied for this choice of auxiliary random variables. The expanded equations are as follows:
\begin{align*}
I(V_1,V_2;Z|V_0) &= I(U_{A},U_{B};Z|U_C) \\
& = I(U_{A\backslash C},U_{B\backslash C};Z_{\backslash C}) \\
& = I(U_{A\backslash C};Z_{,A\backslash C})+ I(U_{B\backslash C};Z_{,B\backslash C}) \\
& = I(V_1;Z|V_0) + I(V_2;Z|V_0),\\
I(V_0, V_1;Y_1) - I(V_0,V_1;Z) &= I(U_A;Y_{1,A}) - I(U_{A};Z_{A}), \\ 
I(V_0, V_1;Y_1) - I(V_0, V_1;Z) &= I(U_B;Y_{1,A}) - I(U_{B};Z_{B}), \\
I(V_1;V_2|V_0) &= I(U_{A\backslash C};U_{B\backslash C}) = 0.
\end{align*}

\noindent \emph{Receivers $Y_1$ and $Y_2$ are less noisy than $Z$}: Recall that in a 2-receiver broadcast channel, a receiver $Y$ is said to be less noisy~\cite{Korner--Marton} than a receiver $Z$ if $I(U;Y) \ge I(U;Z)$ for all $p(u,x)$. In this case, we have
\begin{align*}
C_{\rm S} = \max_{p(x)}\min\{I(X;Y_1) - I(X;Z), I(X;Y_2) - I(X;Z)\}.
\end{align*}
To show achievability, we set $Q = \emptyset$ and $V_0 = V_1 = V_2 = V_3 = X$ in Theorem~\ref{thm:1}. The converse follows similar steps to the converse for Proposition~\ref{prop1} in Subsection~\ref{subsect1} given in Appendix~\ref{appen4} and we omit it here. 

\subsection{2-Receivers, 1-Eavesdropper with Common Message} \label{subsect1}

As a generalization of Theorem~\ref{thm:1}, consider the setting with both common and confidential messages, where we are interested in achieving some equivocation rate for the confidential message rather than asymptotic secrecy. For this setting we can establish the following inner bound on the secrecy capacity region.
\medskip

\begin{theorem} \label{thm:2}
An inner bound to the secrecy capacity region of the 2-receiver, 1-eavesdropper broadcast channel with one common and one confidential messages is given by the set of non-negative rate tuples $(R_0,R_1,R_{e})$ such that {\allowdisplaybreaks
\begin{align*}
R_0 &< I(U;Z), \\
R_0 + R_{1}  &< I(U;Z) +\min\{I(V_0,V_1;Y_1|U) - I(V_1;Z|V_0), I(V_0,V_2;Y_2|U) - I(V_2;Z|V_0)\},\\
R_0 + R_1 &< \min\{I(V_0,V_1;Y_1)-I(V_1;Z|V_0), I(V_0,V_2;Y_2) - I(V_2;Z|V_0)\}, \\
R_e &\le R_1,  \\
R_{e}  &< \min\{I(V_0,V_1;Y_1|U) - I(V_0, V_1;Z|U), I(V_0,V_2;Y_2|U) - I(V_0,V_2;Z|U)\},\\
R_0 + R_e &< \min\{I(V_0,V_1;Y_1)-I(V_1, V_0;Z|U), I(V_0,V_2;Y_2) - I(V_2, V_0;Z|U)\}, \\
R_0  + 2R_e  &<  I(V_0,V_1;Y_1)+I(V_0,V_2;Y_2|U) - I(V_1;V_2|V_0) - 2I(V_0;Z|U), \\
R_0  + 2R_{e} &<  I(V_0,V_2;Y_2) + I(V_0,V_1;Y_1|U) - I(V_1;V_2|V_0) - 2I(V_0;Z|U), \\
R_0 +R_1 + 2R_e  &<  I(V_0,V_2;Y_2|U) - I(V_2;Z|V_0)+ I(V_0,V_1;Y_1)\\
&\quad +I(V_0,V_2;Y_2|U)- I(V_1;V_2|V_0) - 2I(V_0;Z|U), \\
R_0 + R_1 + 2R_{e} &<  I(V_0,V_1;Y_1|U) - I(V_1;Z|V_0)+ I(V_0,V_2;Y_2)\\
&\quad + I(V_0,V_1;Y_1|U)- I(V_1;V_2|V_0) - 2I(V_0;Z|U),
\end{align*}}
for some $p(u,v_0,v_1,v_2,x)=p(u)p(v_0|u)p(v_1,v_2|v_0)p(x|v_0, v_1,v_2)$ such that $I(V_1,V_2;Z|V_0) \le I(V_1;Z|V_0)+ I(V_2;Z|V_0)-I(V_1;V_2|V_0)$.
\end{theorem}
\medskip

Note that if we discard the equivocation rate constraints and set $V_0 = V_1 = V_2 = X$, this inner bound reduces to the straightforward extension of the K\"{o}rner--Marton degraded message set capacity region for the 3 receivers case \cite[Corollary 1]{Nair}. 

If we take $V_0 = V_1 = V_2 = V$ and $Y_1 = Y_2 = Y$, then we obtain the region consisting of all rate pairs $(R_0,R_1)$ such that
\begin{align}
R_0 &< I(U;Z), \\
R_0 + R_{1}  &< I(U;Z) +I(V;Y|U),\nonumber\\
R_0 + R_1 &< I(V;Y), \nonumber\\
R_e &\le R_1,  \nonumber\\
R_{e}  &< I(V;Y|U) - I(V;Z|U),\nonumber\\
R_0 + R_e &< I(V;Y)-I(V;Z|U)
\end{align}
for some $p(u,v,x) = p(u)p(v|u)p(x|v)$.

This region provides an equivalent characterization of the secrecy capacity region of the 2-receiver broadcast channel with confidential messages~\cite{Csiszar}. To see this, note that if we tighten the first inequality to $R_0 \le \min\{I(U;Z), I(U;Y)\}$, the last inequality becomes redundant and the region reduces to the original characterization in \cite{Csiszar}.

\noindent\emph{Proof of Theorem~\ref{thm:2}}:
\medskip

The proof of Theorem \ref{thm:2} involves rate splitting for $R_1(=R_1' + R_1'')$. We first establish an inner bound without rate splitting. The proof with rate splitting is given in Appendix \ref{appen7}. 

\noindent\emph{Codebook generation}: 
Fix $p(u,v_0,v_1,v_2,x)$ and let $R_r\ge 0$ be such that $R_1 - R_e + R_r \ge I(V_0;Z|U) + \d(\e)$. Randomly and independently generate  sequences $u^n(m_0)$, $m_0 \in [1:2^{nR_0}]$, each according to $\prod_{i=1}^np_U(u_i)$. For each $m_0$, randomly and conditionally independently generate  sequences $v_0^n(m_0,m_1, m_r)$, $(m_1,m_r) \in [1:2^{n(R_1+R_r)}]$, each according to $\prod_{i=1}^np_{V_0|U}(v_{0i}|u_i)$. For each $(m_{0},m_1, m_r)$, generate  sequences $v_1^n(m_{0},m_1, m_r, t_1)$, $t_1 \in [1:2^{nT_1}]$, each according to $\prod_{i=1}^n p_{V_1|V_0}(v_{1i}|v_{0i})$, and partition the set $[1:2^{nT_1}]$ into $2^{n\Rt_1}$ equal size bins $\Bc(m_0, m_1, m_r, l_1)$. Similarly, for each $(m_{0},m_1, m_r)$, randomly generate  sequences $v_2^n(m_{0},m_1, m_r, t_2)$, $t_2 \in [1:2^{nT_2}]$ each according to $\prod_{i=1}^np_{V_2|V_0} (v_{2i}|v_{0i})$ and partition the set $[1:2^{nT_2}]$ into $2^{n\Rt_2}$ bins $\Bc(m_0, m_1, m_r, l_2)$. Finally, for each product bin $\Bc(l_{1})\times \Bc(l_{2}) $, find a jointly typical sequence pair $(v_1^n(m_{0},m_1, m_r, t_1(l_1)),v_2^n(m_{0},m_1, m_r, t_2(l_2))$. If there is more than 1 pair, we randomly and uniformly pick a pair from the set of jointly typical pairs. As before, the probability of error approaches zero as $n \to \infty$ if  
\begin{align*}
\Rt_1 + \Rt_2 < T_1 + T_2 - I(V_1;V_2|V_0) - \d(\e).
\end{align*}

\noindent\emph{Encoding}:
To send a message pair $(m_0,m_1)$, the encoder first chooses a random index $m_r \in [1:2^{nR_r}]$ and then the sequence pair $(u^n(m_0),v_0^n(m_1,m_r, m_0))$. It then randomly chooses a product bin indices $(L_1,L_2)$ and selects the jointly typical sequence pair \\ $(v_1^n(m_0, m_1, m_r, t_1(L_1)), v_2^n(m_0,m_1, m_r,t_2(L_2))$ in it. Finally, it generates a codeword $X^n$ at random\\ according to $\prod_{i=1}^np_{X|V_0,V_1.V_2} (x_i|v_{0i}, v_{1i},v_{2i})$. 
\medskip

\noindent\emph{Decoding and analysis of the probability of error}:
Receiver $Y_1$ finds $(m_0,m_1)$ indirectly by looking for the unique $(m_0,\lh_0)$ such that $(u^n(m_0), v_0^n(m_0,\lh_0), v_1^n(m_0, \lh_0, l_1)) \in \aep$ for some $l_1 \in [1:2^{nT_1}]$. Similarly, receiver $Y_2$ finds $(m_0,m_1)$ indirectly by looking for the unique $(m_0,\lh_0)$ such that $(u^n(m_0), v_0^n(m_0,\lh_0), v_1^n(m_0, \lh_0, l_2)) \in \aep$ for some $l_2 \in [1:2^{nT_2}]$. Receiver $Z$ finds $m_0$ directly by decoding $U$. These steps succeed with probability of error approaching zero as $n \to \infty$ if
\begin{align*}
R_0 + R_1 + T_1 +R_r &< I(V_0, V_1;Y_1)-\d(\e),\\
R_{1} + T_1 +R_r &< I(V_0, V_1;Y_1|U)-\d(\e), \\
R_0 + R_1 + T_2 +R_r&< I(V_0, V_1;Y_2)-\d(\e), \\
R_{1} + T_2 +R_r&< I(V_0, V_1;Y_2|U)-\d(\e),\\
R_0 &< I(U;Z).
\end{align*}
\noindent\emph{Analysis of equivocation rate}: We consider the equivocation rate averaged over codes. We will show that a part of the message $M_{1p}$ can be kept asymptotically secret from the eavesdropper as long as rate constraints on $R_e$ and $R_1$ are satisfied. Let $R_1 = R_{1p} + R_{1c}$ and $R_e = R_{1p}$.  
\begin{align}
H(M_{1}|Z^n, \Cc) & \geq H(M_{1p}|Z^n, M_0, \Cc) \nonumber \\
&= H(M_{1p}) - I(M_{1p};Z^n| M_0, \Cc) \label{eqn:a}\\
& \stackrel{(a)}{\geq} H(M_{1p}) - 3n\d(\e) \nonumber \\
& = n(R_{1} - I(V_0;Z|U)) -3n\d(\e). \nonumber
\end{align}
This implies that $R_{e} \leq R_1 - I(V_0;Z|U) -3\d(\e)$
is achievable.

To prove step $(a)$, consider 
\begin{align*}
I(M_{1p};Z^n| M_0, \Cc) & = I(T_1(L_1), T_2(L_2), M_{1p}, M_{1c}, M_r; Z^n| M_0, \Cc) - I(T_1(L_1), T_2(L_2), M_{1c}, M_r; Z^n|M_{1p}, M_0,\Cc) \\
& \stackrel{(b)}{\le} I(V_0^n, V_1^n, V_2^n; Z^n| M_0, \Cc) - I(M_{1c}, M_r; Z^n |M_{1p}, M_0, \Cc)- I(T_1(L_1), T_2(L_2); Z^n|M_{1},  M_0, M_r, \Cc) \\
& \stackrel{(c)}{\le} I(V_0^n, V_1^n, V_2^n; Z^n| U^n, \Cc) - I(M_{1c}, M_r; Z^n |M_{1p}, M_0, \Cc) - I(T_1(L_1), T_2(L_2); Z^n|M_{1},  M_0, M_r, \Cc) \\
& \leq nI(V_0, V_1, V_2; Z|U) +n\d(\e)-  H(M_{1c}, M_r |M_{1p}, U^n, \Cc) + H(M_{1c}, M_r |M_{1p}, M_0, Z^n, \Cc)\\
&\qquad - I(T_1(L_1), T_2(L_2); Z^n|M_1,M_0, M_r, \Cc) \\
& \le  nI(V_0, V_1, V_2; Z|U) + n\d(\e) -  n(R_{1} - R_e + R_r) + H(M_{1c}, M_r |M_{1p}, M_0, Z^n, \Cc)\\
&\qquad - I(T_1(L_1), T_2(L_2); Z^n|M_1,M_0,M_r, \Cc),
\end{align*}
where $(b)$ follows by the data processing inequality and $(c)$ follows by the observation that $U^n$ is a function of $(\Cc,M_0)$ and $(\Cc,M_0)\to (\Cc,U^n,V^n)\to Z^n$. Following the analysis of the equivocation rate terms in Theorem~\ref{thm:1} and using Lemma 1, the remaining terms can be bounded by
{\allowdisplaybreaks
\begin{align*}
H(M_{1c}, M_r |M_{1p}, M_0, Z^n, \Cc) &\le H(M_{1c}, M_r |M_{1p}, U^n, Z^n)\\
&\le   n(R_{1} - R_e + R_r)  -nI(V_0;Z|U) + n\d(\e), \\
I(T_1(L_1), T_2(L_2); Z^n|M_1,M_0, M_r, \Cc) & = H(T_1(L_1), T_2(L_2)|M_1,M_0, M_r, \Cc) - H(T_1(L_1), T_2(L_2)|M_1,M_0, M_r,\Cc, Z^n)\\
& = n(\Rt_1 + \Rt_2) - H(T_1(L_1), T_2(L_2)|M_1,M_0, M_r, \Cc, Z^n)\\
& \stackrel{(a)}{=} n(\Rt_1 + \Rt_2) - H(T_1(L_1), T_2(L_2)|M_r, M_1,M_0,V_0^n, \Cc, Z^n)\\
& \ge n(\Rt_1 + \Rt_2) - H(T_1(L_1), T_2(L_2)|V_0^n, Z^n)\\
& \ge n(\Rt_1 + \Rt_2 - T_1 - T_2) + n(I(V_1;Z|V_0) + I(V_2;Z|V_0)) - 2n\d(\e),
\end{align*}
if  $T_1 \ge I(V_1;Z|V_0) + \d(\e)$, and $T_2 \ge I(V_2;Z|V_0) + \d(\e)$. Step $(a)$ follows from the observation that $V_0^n$ is a function of $(\Cc,M_0,M_1)$.}

Thus, we have
\begin{align*}
I(M_{1p};Z^n| M_0, \Cc) &\le I(V_1,V_2;Z|V_0) -I(V_1;Z|V_0) - I(V_2;Z|V_0)  + n(T_1 +T_2 - \Rt_1 - \Rt_2) + 4n\d(\e).
\end{align*}
Hence, $I(M_{1p};Z^n| M_0, \Cc) \le 4n\d(\e)$ if
\begin{align*}
I(V_1;V_2;Z|V_0) + T_1 + T_2 -\Rt_1 - \Rt_2 - I(V_1;Z|V_0) - I(V_2;Z|V_0)) \le 0.
\end{align*}

Substituting back into~(\ref{eqn:a}) shows that
\begin{align*}
H(M_{1}|Z^n, \Cc) & \ge n(R_1-I(V_0;Z|U) - 4n\d(\e).
\end{align*}
The equivocation rate constraints on the rates are 
\begin{align*}
R_e &\le R_1, \\
R_r &\ge 0, \\
R_1 - R_e + R_r &\ge I(V_0;Z|U), \\
T_1 &\ge I(V_1;Z|V_0), \\
T_2 &\ge I(V_2;Z|V_0).
\end{align*}
Using Fourier-Motzkin elimination then gives us an inner bound for the case without rate splitting. The proof with rate splitting on $R_1$ is given in Appendix \ref{appen7}.

\subsection*{Special Case:}
We show that the inner bound in Theorem~\ref{thm:2} is tight when both $Y_1$ and $Y_2$ are less noisy than $Z$. 
\medskip
\begin{proposition} \label{prop1}
When both $Y_1$ and $Y_2$ are less noisy than $Z$, the 2-receiver, 1-eavesdropper secrecy capacity region is given by the set of $(R_0,R_1, R_e)$ tuples such that
\begin{align*}
R_0 &\leq I(U;Z), \\
R_1 &\leq \min \{I(X;Y_1|U), I(X;Y_2|U)\}, \\
R_e & \leq [\min\{R_1, I(X;Y_1|U) - I(X;Z|U), I(X;Y_2|U) - I(X;Z|U)\}]^+
\end{align*}
for some $p(u,x)$. 
\end{proposition}
\medskip
Achievability follows by setting $V_0=V_1=V_2=X$ in Theorem~\ref{thm:2} and using the fact that $Y_1$ and $Y_2$ are less noisy than $Z$, which allows us to assume without loss of generality that $R_0 \le \min\{I(U;Z), I(U;Y_1), I(U;Y_2)\}$. The set of inequalities then reduce to
\begin{align*}
R_0 &< I(U;Z), \\
R_0 + R_{1}  &< I(U;Z) +\min\{I(X;Y_1|U), I(X;Y_2|U) \},\\
R_e &\le R_1,  \\
R_{e}  &< \min\{I(X;Y_1|U) - I(X;Z|U), I(X;Y_2|U) - I(X;Z|U)\}.
\end{align*}
Since the region in Proposition~\ref{prop1} is a subset of the above region, we have established the achievability part of the proof. Achievability in this case, however, is a straightforward extension of Csisz\'{a}r and K\"{o}rner and does not require Marton coding.
For the converse, we use the identification $U_i = (M_0, Z^{i-1})$. With this identification, the $R_0$ inequality follows trivially. The $R_1$ and $R_e$ inequalities follow from standard methods and a technique in~\cite[Proposition 11]{Nair}. The details are given in Appendix~\ref{appen4}.

\section{1-receiver, 2-eavesdroppers wiretap channel} \label{sect:5}
We now consider the case where the confidential message $M_1$ is to be sent only to $Y_1$ and kept hidden from the eavesdroppers $Z_2$ and $Z_3$. All three receivers $Y_1, Z_2, Z_3$ require a common message $M_0$. For simplicity, we only consider the special case of multilevel broadcast channel~\cite{bzt}, where $p(y_1, z_2, z_3|x) = p(y_1, z_3|x)p(z_2|y_1)$. In \cite{Nair}, it was shown that the capacity region (without secrecy) is the set of rate pairs $(R_0,R_1)$ such that
\begin{align*}
R_0 &< \min \{I(U; Z_2), I(U_3; Z_3)\}, \\
R_1 &< I(X;Y_1|U), \\
R_0 + R_1 &< I(U_3;Z_3) + I(X;Y_1|U_3) 
\end{align*}
for some $p(u)p(u_3|u)p(x|u_3)$. We extend this result to obtain inner and outer bounds on the secrecy capacity region.\\

\begin{proposition} \label{prop2}
An inner bound to the secrecy capacity region of the 1-receiver, 2-eavesdropper multilevel broadcast channel with common and confidential messages is is given by the set of rate tuples $(R_0,R_1, R_{e2}, R_{e3})$ such that
{\allowdisplaybreaks
\begin{align*}
R_0 &< \min \{I(U; Z_2), I(U_3; Z_3)\}, \\
R_1 &< I(V;Y_1|U), \\
R_0 + R_1 &< I(U_3;Z_3) + I(V;Y_1|U_3), \\
R_{e2} &\le \min \{R_1, I(V;Y_1|U) - I(V;Z_2|U)\}, \\
R_{e2} &\le [I(U_3;Z_3)-R_0 - I(U_3;Z_2|U)]^+ + I(V;Y_1|U_3) - I(V;Z_2|U_3), \\
R_{e3} &\leq \min \{R_1, [I(V;Y_1|U_3) - I(V;Z_3|U_3)]^+ \}, \\
R_{e2} + R_{e3} &\le R_1 +  I(V;Y_1|U_3) - I(V;Z_2|U_3),
\end{align*}
for some} $p(u,u_3,v,x) = p(u)p(u_3|u)p(v|u_3)p(x|v)$. 
\end{proposition}
\medskip

It can be shown that setting $Y_1 = Z_2 = Y$ and $Z_3 = Z$ gives an alternative characterization of the secrecy capacity of the broadcast channel with confidential messages.  

\emph{Proof of achievability}: 
We break down the proof of Proposition \ref{prop2} into four cases and give the analysis of the first case in detail. The analyses  for the rest of the cases are similar and we therefore we only provide a sketch in Appendix \ref{appen6}. Furthermore, in all cases, we assume that $R_1 \ge \min\{I(V;Y_1|U_3) - I(V;Z_2|U_3), [I(V;Y_1|U_3) - I(V;Z_3|U_3)]^+\}$. It is easy to see from our proof that if this inequality does not hold, then we achieve equivocation rates of $R_{e2} = R_{e3} = R_1$ for any rate pair$(R_0, R_1)$ satisfying the inequalities in the proposition. The four cases are:
\begin{itemize}
\item Case 1: $I(U_3;Z_3) - R_0 - I(U_3;Z_2|U) \ge 0$, $I(V;Y_1|U_3) - I(V;Z_2|U_3) \le I(V;Y_1|U_3) - I(V;Z_3|U_3)$ and $R_{e3} \ge I(V;Y_1|U_3) - I(V;Z_2|U_3)$;\\
\item Case 2: $I(U_3;Z_3) - R_0 - I(U_3;Z_2|U) \ge 0$, $I(V;Y_1|U_3) - I(V;Z_2|U_3) \le I(V;Y_1|U_3) - I(V;Z_3|U_3)$ and $R_{e3} \le I(V;Y_1|U_3) - I(V;Z_2|U_3)$;\\
\item Case 3: $I(U_3;Z_3) - R_0 - I(U_3;Z_2|U) \ge 0$, $I(V;Y_1|U_3) - I(V;Z_2|U_3) \ge I(V;Y_1|U_3) - I(V;Z_3|U_3)$. In this case, since we consider only the case of $R_1 \ge I(V;Y_1|U_3) - I(V;Z_3|U_3)$, we will see that an equivocation rate of $R_{e3} = I(V;Y_1|U_3) - I(V;Z_3|U_3)$ can be achieved;\\
\item Case 4: $I(U_3;Z_3) - R_0 - I(U_3;Z_2|U) \le 0$.
\end{itemize}

Now, consider Case 1, where $I(U_3;Z_3) - R_0 - I(U_3;Z_2|U) \ge 0$, $I(V;Y_1|U_3) - I(V;Z_2|U_3) \le I(V;Y_1|U_3) - I(V;Z_3|U_3)$ and $R_{e3} \ge I(V;Y_1|U_3) - I(V;Z_2|U_3)$. 

\noindent \emph{Codebook generation}: Fix $p(u,u_3,v,x) = p(u)p(u_3|u)p(v|u_3)p(x|v)$. Let $R_1 = R_{10}^o + R_{10}^s + R_{11}' + R_{11}'' + R_{11}^o$. Let $R_0^r\ge 0$ and $R_1^r\ge 0$ be the randomization rates introduced by the encoder. These are not part of the message rate. Let $\Rt_{10} = R_{10}^o + R_{10}^s + R_0^r$ and $\Rt_{11} =R_{11}' + R_{11}'' + R_{11}^o + R_1^r $.

Randomly and independently generate  sequences $u^n(m_0)$, $m_0 \in [1:2^{nR_0}]$, each according to $\prod_{i=1}^np_U(u_i)$. For each $m_0$, randomly and conditionally independently generate sequences $u_3^n(m_0,l_0)$, $l_0 \in [1:2^{n\Rt_{10}}]$, each according to $\prod_{i=1}^np_{U_3|U}(u_{3i}|u_i)$. For each $(m_{0},l_0)$, randomly and conditionally independently generate sequences $v^n(m_{0},l_0,l_1)$, $l_1 \in [1:2^{n\Rt_{11}}]$, each according to $\prod_{i=1}^n p_{V|U_3}(v_{i}|u_{3i})$.
\medskip

\noindent\emph{Encoding}: To send a message $(m_0,m_1)$, we split $m_1$ into sub-messages with the corresponding rates given in the codebook generation step and generate the randomization messages $(m_{10}^r,m_{11}^r)$ uniformly at random  from the set $[1:2^{nR_{0}^r}] \times [1:2^{nR_1^r}]$. We then select the sequence $v^n(m_0,l_0,l_1)$ corresponding to $(m_0,m_1, m_{10}^r, m_{11}^r)$ and send $X^n$ generated according to $\prod_{i=1}^np_{X|V}(x_i|v_i(l_1,l_0,m_0))$.
\medskip

\noindent \emph{Decoding and analysis of the probability of error}: 
Receiver $Y_1$ finds $(m_0,m_1)$ by decoding $(U,U_3,V)$, $Z_2$ finds $m_0$ by decoding $U$, and $Z_3$ finds $m_0$ indirectly through $(U, U_3)$. The probability of error goes to zero as $n\to \infty$ if
\begin{align*}
R_0 \le I(U;Z_2), \\
R_{0} + R_{10}^{o} + R_{0}^r+R_{10}^s &< I(U_3;Z_3) - \d(\e), \\
R_{10}^s + R_{10}^o + R_0^r &< I(U_3;Y_1|U) - \d(\e), \\
R_{11}' + R_{11}'' + R_{11}^o + R_{1}^r &< I(V;Y_1|U_3) -\d(\e).
\end{align*}


\noindent \emph{Analysis of equivocation rates}: We show that the following equivocation rates are achievable.
\begin{align*}
R_{e2}& = R_{10}^s + R_{11}' - \d(\e),  \\
R_{e3} &=R_{11}' +R_{11}'' - \d(\e).
\end{align*}

It is straightforward to show that the stated equivocation rate $R_{e3}$ is achievable if
\begin{align*}
R_1^r + R_{11}^o &> I(V;Z_3|U_3) + \d(\e).
\end{align*}

The analysis of the $H(M_1|Z_2^n, \Cc)$ term is slightly more involved. Consider
\begin{align*}
I(M_{10}^s, M_{11}'; Z_2^n| M_0, \Cc) &=  I(L_0, L_1; Z_2^n|\Cc, M_0) - I(L_0, L_1; Z_2^n|\Cc, M_0, M_{10}^s, M_{11}') \\
&\le I(V^n;Z^n_2|\Cc, U^n) - I(L_0; Z_2^n|\Cc, M_0, M_{10}^s, M_{11}') - I(L_1; Z_2^n|\Cc, M_0, L_0, M_{11}') \\
& \le \sum_{i=1}^nI(V;Z_2|U) - I(L_0; Z_2^n|\Cc, M_0, M_{10}^s, M_{11}') - I(L_1; Z_2^n|\Cc, M_0, L_0, M_{11}'). 
\end{align*}

Now consider the second and third terms. We have
\begin{align*}
I(L_0; Z_2^n|\Cc, M_0, M_{10}^s, M_{11}') &= H(L_0|\Cc, M_0, M_{10}^s, M_{11}') - H(L_0|\Cc, M_0, M_{10}^s, M_{11}', Z_2^n, U^n) \\
 &\ge n(\Rt_{10} - R_{10}^s) - H(L_0|\Cc,M_{10}^s, Z_2^n, U^n) \\
 &\ge n(I(U_3;Z_2|U) - \d(\e)).
\end{align*}
The last step follows from Lemma \ref{lem1}, which holds if 
\begin{align*}
\Rt_{10} - R_{10}^s &= R_{10}^o + R_{0}^r \\
& \ge I(U_3;Z_2|U) + \d(\e).
\end{align*}

For the third term, we have 
\begin{align*}
I(L_1; Z_2^n|\Cc, M_0, L_0, M_{11}') & = H(L_1|\Cc, M_0, L_0, M_{11}') - H(L_1|\Cc, M_0, L_0, M_{11}', Z_2^n, U^n) \\
& \ge n(\Rt_{11} - R_{11}') - H(L_1|\Cc, M_{11}', Z_2^n, U^n) \\
& \ge  n(\Rt_{11} - R_{11}') - n(\Rt_{11} - R_{11}' - I(V;Z_2|U_3) + \d(\e)). 
\end{align*}
In the last step, we again apply Lemma \ref{lem1}, which holds if
\begin{align*}
R_{11}'' + R_{11}^o + R_{1}^r \ge I(V;Z_|U_3) + \d(\e).
\end{align*}

In summary, the inequalities for Case 1 are as follows:

\emph{Decoding Constraints}: (with $R_0 \le I(U;Z_2)$ omitted since this inequality appears in the final rate-equivocation region and does not contain the auxiliary rates to be eliminated.)
\begin{align*}
R_{0} + R_{10}^{o} + R_{0}^r+R_{10}^s &< I(U_3;Z_3), \\
R_{10}^s + R_{10}^o + R_0^r &< I(U_3;Y_1|U), \\
R_{11}' + R_{11}'' + R_{11}^o + R_{1}^r &< I(V;Y_1|U_3).
\end{align*}

\emph{Equivocation rate constraints}:
\begin{align*}
R_{10}^o + R_o^r &> I(U_3;Z_2|U), \\
R_{11}'' + R_1^r + R_{11}^o &> I(V;Z_2|U_3), \\
R_1^r + R_{11}^o &> I(V;Z_3|U_3).
\end{align*}

\emph{Greater than or equal to zero constraints}:
\begin{align*}
R_{10}^o , R_0^o, R_{11}', R_{11}'',R_1^n, R_0^r \ge 0. 
\end{align*}

\emph{Equality constraints}:
\begin{align*}
R_1 &= R_{10}^o + R_{10}^s + R_{11}' + R_{11}'' + R_{11}^o, \\
R_{e2}& = R_{10}^s + R_{11}',  \\
R_{e3} &=R_{11}' +R_{11}''.
\end{align*}

Applying Fourier-Motzkin elimination yields the rate-equivocation region for Case one. Sketch of achievability for the other cases are given in Appendix \ref{appen6}.

We now establish an outer bound and use it to show that the inner bound in Proposition \ref{prop2} is tight in several special cases. In contrast to the case with no secrecy constraint~\cite{Nair}, the assumption of a stochastic encoder makes it difficult to match our inner and outer bounds in general. 
\medskip

\begin{proposition} \label{prop3}
An outer bound on the secrecy capacity of the multilevel 3-receiver broadcast channel with one common and one confidential messages is given by the set of rate tuples $(R_0,R_1,R_{e2},R_{e3})$ such that
{\allowdisplaybreaks
\begin{align*}
R_0 & \leq \min\{I(U;Z_2), I(U_3;Z_3)\},\\
R_1 &\leq I(V;Y_1|U), \\
R_0 + R_1 & \leq I(U_3;Z_3) + I(V;Y_1|U_3), \\
R_{e2} &\leq  I(X;Y_1|U) - I(X;Z_2|U), \\
R_{e2} &\leq [I(U_3;Z_3)-R_0 - I(U_3;Z_2|U)]^+ + I(X;Y_1|U_3) - I(X;Z_2|U_3), \\
R_{e3} &\leq [I(V;Y_1|U_3) - I(V;Z_3|U_3)]^+
\end{align*}
}for some $p(u,u_3,v, x) = p(u)p(u_3|u) p(v|u_3) p(x|v)$. 
\end{proposition}
Proof of this Proposition uses a combination of standard converse techniques from \cite{Abbas}, \cite{Abbas3}, and \cite{Csiszar} and given in Appendix \ref{appen5}.
\medskip

\noindent\emph{Remark 5.1}: As we can see in the inequalities governing $R_{e2}$ in both the inner and outer bounds, there is a tradeoff between the common message rate and the equivocation rate at receiver $Z_2$. A higher common message rate limits the number of codewords that can be generated to confuse the eavesdropper.

\subsection*{Special Cases}
Using Propositions \ref{prop2} and \ref{prop3}, we can establish the secrecy capacity region for the following special cases.

\noindent\emph{$Y_1$ more capable than $Z_3$ and $Z_3$ more capable than $Z_2$}: If $Y_1$ is \emph{more capable}~\cite{Abbas} than $Z_3$ and $Z_3$ is more capable than $Z_2$, the capacity region is given by:
\begin{align*}
R_0 & \leq \min\{I(U;Y_2), I(U_3;Z_3)\},\\
R_1 &\leq I(X;Y_1|U), \\
R_0 + R_1 & \leq I(U_3;Z_3) + I(X;Y_1|U_3), \\
R_{e2} &\leq  I(X;Y_1|U) - I(X;Z_2|U), \\
R_{e2} &\le [I(U_3;Z_3)-R_0 - I(U_3;Z_2|U)]^+ + I(X;Y_1|U_3) - I(X;Z_2|U_3),\\
R_{e3} &\leq  I(X;Y_1|U_3) - I(X;Z_3|U_3)
\end{align*}
for some $p(u,u_3, x) = p(u)p(u_3|u)p(x|u_3)$.

Achievability follows directly from setting $V =X$ and observing that since $Z_3$ is more capable than $Z_2$, the inequality $R_{e2} + R_{e3} \le R_1 +  I(X;Y_1|U_3) - I(X;Z_2|U_3)$ is redundant since $I(X;Y_1|U_3) - I(X;Z_2|U_3) \ge I(X;Y_1|U_3) - I(X;Z_3|U_3)$ from the more capable condition. For the converse, observe that since $Y_1$ is more capable than $Z_3$, we have
\begin{align*}
I(V;Y_1|U_3) - I(V;Z_3|U_3) &= I(V,X;Y_1|U_3) - I(V,X;Z_3|U_3)  -I(X;Y_1|V) + I(X;Z_3|V) \\
& \leq I(X;Y_1|U_3) - I(X;Z_3|U_3). 
\end{align*}

\emph{One eavesdropper}: Here, we consider the two scenarios where either $Z_2$ or $Z_3$ is an eavesdropper and the other receiver is neutral, i.e., there is no constraint on its equivocation rate, but it still decodes a common message. The secrecy capacity regions for these two scenarios are as follows.
\medskip

\noindent\emph{$Z_3$ is neutral}: The secrecy capacity region is the set of rate tuples $(R_0, R_1, R_{e2})$ such that
\begin{align*}
R_0 & \leq \min\{I(U;Y_2), I(U_3;Z_3)\},\\
R_1 &\leq I(X;Y_1|U), \\
R_0 + R_1 & \leq I(U_3;Z_3) + I(X;Y_1|U_3), \\
R_{e2} &\leq  I(X;Y_1|U) - I(X;Z_2|U), \\
R_{e2} &\le  [I(U_3;Z_3)-R_0 - I(U_3;Z_2|U)]^+ + I(X;Y_1|U_3) - I(X;Z_2|U_3)
\end{align*}
for some $p(u,u_3, x) = p(u)p(u_3|u)p(x|u_3)$.
\medskip

\noindent\emph{$Z_2$ is neutral}: The secrecy capacity region is the set of rate tuples $(R_0, R_1, R_{e3})$ such that
\begin{align*}
R_0 & \leq \min\{I(U;Z_2), I(U_3;Z_3)\},\\
R_1 &\leq I(V;Y_1|U), \\
R_0 + R_1 & \leq I(U_3;Z_3) + I(V;Y_1|U_3), \\
R_{e3} &\leq  [I(V;Y_1|U_3) - I(V;Z_3|U_3)]^+
\end{align*}
for some $p(u,u_3, v,x) = p(u)p(u_3|u)p(v|u_3)p(x|v)$.

\section{Conclusion}
We presented inner and outer bounds on the secrecy capacity region of the 3-receiver broadcast channel with common and confidential messages that are strictly larger than straightforward extensions of the Csisz\'ar--K{\"o}rner 2-receiver region. We considered the 2-receiver, 1-eavesdropper and the 1-receiver, 2-eavesdroppers cases. For the first case, we showed that additional superposition encoding, whereby a codeword is picked at random from a pre-generated codebook can increase the achievable rate by allowing the legitimate receiver to indirectly decode the message without sacrificing secrecy. A general lower bound on the secrecy capacity is then obtained by combining superposition encoding and indirect decoding with Marton coding. This lower bound is shown to be tight for the reversely degraded product channel and when both $Y_1$ and $Y_2$ are less noisy than the eavesdropper. The lower bound was generalized in Theorem~\ref{thm:2} to obtain an inner bound on the secrecy capacity region for the 2-receiver, 1 eavesdropper case. For the case where both $Y_1$ and $Y_2$ are less noisy than the eavesdropper, we again show that our inner bound gives the secrecy capacity region. 

We then established inner and outer bounds on the secrecy capacity region for the 1-receiver, 2-eavesdroppers multilevel wiretap channel. The inner bound and outer bounds are shown to be tight for several special cases.
In the results for both setups, we observe a tradeoff between the common message rate and the eavesdropper equivocation rates. A higher common message rate limits the number of codewords that can be generated to confuse the eavesdroppers about the confidential message. In addition, in the second setup, a higher common message rate can potentially reduce the equivocation rate of one eavesdropper while leaving the equivocation rate at the other eavesdropper unchanged.

\section*{Acknowledgment}
The authors would like to thank Chandra Nair and Han-I Su for helpful comments and the anonymous reviewers for many insightful remarks that helped greatly improve the paper.

\bibliographystyle{IEEEtran}
\bibliography{secrecy}

\appendices \section{Proof of Lemma~\ref{lem1}} \label{appen1}

First, define $N(U^n,Z^n)= |\{k\in [1:2^{nS}]: (U^n, V^n(k),Z^n)\in \aep\}|$. Next, we define the following ``error'' events. Let $E_1(U^n, Z^n)=1$ if $\{N(U^n, Z^n) \ge (1+\d_1(\e))2^{n(S - I(V;Z|U)+\d(\e))}\}$ and $E_1=0$ otherwise. Let $E = 0$ if $(U^n, V^n(L),Z^n)\in \aep$ and $E_1(U^n,Z^n,L) = 0$, and $E =1$ otherwise. We now show that if $S \ge I(V;Z|U) + \d(\e)$, then $\P\{E = 1\} \to 0$ as $n \to \infty$. By the union of events bound,
\begin{align*}
\P\{E = 1\} &\le \P\{(U^n,V^n(L),Z^n)\notin \aep\} + \P\{E_1(U^n,Z^n,L) = 1\}.
\end{align*}
The first term tends to zero as $n \to \infty$ by assumption. The second term is bounded as follows
\begin{align*}
\P\{E_1(U^n,Z^n) = 1\} & = \sum_{u^n \in \aep} p(u^n)\P\{(E_1(U^n,Z^n) = 1)|U^n = u^n\} \\
& = \sum_{u^n \in \aep}\sum_{z^n \in \aep(Z|U)} p(u^n)\P\{(E_1(u^n,Z^n) = 1)\cap (Z^n = z^n)|U^n = u^n\} \\
& = \sum_{u^n \in \aep}\sum_{z^n \in \aep(Z|U)} p(u^n)\P\{(E_1(u^n,z^n) = 1)\cap (Z^n = z^n)|U^n = u^n\} \\
& \le \sum_{u^n \in \aep}p(u^n)\sum_{z^n \in \aep(Z|U)}\P\{(E_1(u^n,z^n) = 1)|U^n = u^n\}.
\end{align*}
Now, $\P \{E_1(u^n,z^n) = 1|U^n = u^n \} = \P \{N(u^n,z^n)\ge (1+\d_1(\e))2^{n(S - I(V;Z|U)+\d(\e))} |U^n = u^n\}$. Define $X_k=1$ if $(u^n, V^n(k),z^n)\in \aep$ and $0$, otherwise. We note that $X_k$, $k \in [1:2^{nS}]$, are i.i.d. Bernoulli $p$ random variables, where $2^{-n(I(V;Z|U) + \d(\e))} \le p \le 2^{-n(I(V;Z|U) - \d(\e))}$. We have
\begin{align*}
&\P\{N(u^n,z^n)\ge (1+\d_1(\e))2^{n(S - I(V;Z|U)+\d(\e))} |U^n = u^n\} \\
&\le \P\left\{\sum_{k=1}^{2^{nS}} X_k \ge (1+\d_1(\e))2^{nS}p |U^n = u^n \right\}. 
\end{align*}
Applying the Chernoff Bound (e.g., see \cite[Appendix B]{El-Gamal--Kim2010}), we have
\begin{align*}
\P\left\{\sum_{k=1}^{2^{nS}} X_k \ge (1+\d_1(\e))2^{nS}p |U^n = u^n \right\} &\le \exp(-2^{nS}p\d_1^2(\e)/4) \\
& \le \exp(-2^{n(S - I(V;Z|U) - \d(\e))}\d_1^2(\e)/4).
\end{align*}

Hence, 
\begin{align*}
\P\{E_1(U^n,Z^n) = 1\} & \le \sum_{u^n \in \aep}p(u^n)\sum_{z^n \in \aep(Z|U)} \exp(-2^{n(S - I(V;Z|U) - \d(\e))}\d_1^2(\e)/4) \\
& \le 2^{n\log|\Zc|} \exp(-2^{n(S - I(V;Z|U) - \d(\e))}\d_1^2(\e)/4),
\end{align*}
which tends to zero as $n \to \infty$ if $S > I(V;Z|U) + \d(\e)$. 

We are now ready to bound $H(L|\Cc, Z^n,U^n)$. Consider
\begin{align*}
H(L, E|\Cc, U^n,Z^n) & \le 1+ \P\{E = 1\}H(L|\Cc, E = 1, U^n,Z^n) + \P\{E = 0\}H(L|\Cc, E = 0, U^n,Z^n) \\
& \le 1 + \P\{E = 1\}nS + \log ((1+\d_1(\e))2^{n(S - I(V;Z|U)+\d(\e))}) \\
& \le n(S -I(V;Z|U) + \d'(\e)).
\end{align*} 
This completes the proof of the lemma.
\section{Evaluation for example} \label{appen2}

We first give an upper bound for the extended Csisz\'{a}r--K\"{o}rner lower bound.

\emph{Fact}: The extended Csisz\'{a}r and K\"{o}rner lower bound in (\ref{eqn:2}) for the channel shown in Figure~\ref{fig1} is upper bounded by 
\begin{align*}
R_{\rm CK} &\leq \min  \{I(X_1;Y_{11}) - I(X_1;Z_{1})+ I(V_2;Y_{12}|Q_2) - I(V_2;Z_{2}|Q_2),  I(X_1;Y_{21})\\
& \qquad \qquad  - I(X_1;Z_{1})- I(V_2;Z_{2}|Q_2)\}.
\end{align*}
for some $p(x_1)p(q_2, v_2)p(x_2|v_2)$.

\begin{proof}
From (\ref{eqn:2}), we have
\begin{align*}
R \le \max_{p(q)p(v|q)p(x|v)} \min\{I(V;Y_1|Q) - I(V;Z|Q), I(V;Y_2|Q) - I(V;Z|Q)\}.
\end{align*}
Consider the first bound for $R_{\rm CK}$.
\begin{align*}
I(V;Y_1|Q) - I(V;Z|Q) &= I(V;Y_{11}, Y_{12}|Q) - I(V;Z_1|Q) - I(V;Z_2|Q,Z_1) \\
& \le I(V;Y_{11}, Y_{12}, Z_{1}|Q) - I(V;Z_1|Q) - I(V;Z_2|Q,Z_1) \\
& = I(V;Y_{11}, Y_{12}|Q, Z_1)  - I(V;Z_2|Q,Z_1) \\
& = I(V;Y_{11}|Q, Z_1, Y_{12}) + I(V;Y_{12}|Q, Z_1)  - I(V;Z_2|Q,Z_1) \\
& = I(V;Y_{11}, Z_1|Q, Y_{12}) - I(V;Z_1 | Q, Y_{12}) + I(V;Y_{12}|Q, Z_1)  - I(V;Z_2|Q,Z_1) \\
& \stackrel{(a)}{=} I(V;Y_{11}|Q, Y_{12}) - I(V;Z_1 | Q, Y_{12}) + I(V;Y_{12}|Q, Z_1)  - I(V;Z_2|Q,Z_1) \\
& \le I(V';Y_{11}|Q) - I(V';Z_1 | Q) + I(V;Y_{12}|Q, Z_1)  - I(V;Z_2|Q,Z_1).
\end{align*}
$(a)$ follows from the structure of the channel which gives the Markov condition $(Q,Y_{12}, V)-Y_{11}- Z_1$. The last step follows from defining $V' = (V, Y_{12})$ and the fact that $Z_1$ is a degraded version of $Y_{11}$.

Consider now the second bound.
\begin{align*}
R_{\rm CK} &\le I(V;Y_2|Q) - I(V;Z|Q) \\
& =  I(V;Y_{21}|Q) - I(V;Z_1|Q) - I(V;Z_2|Q, Z_1) \\
& \le  I(V';Y_{21}|Q) - I(V';Z_1|Q) - I(V;Z_2|Q, Z_1).
\end{align*}
Combining the bounds, we have
\begin{align}
R_{\rm CK} \le \max_{p(q,v, v', x_1, x_2)} \min & \left\{ I(V';Y_{11}|Q) - I(V';Z_1 | Q) + I(V;Y_{12}|Q, Z_1)  - I(V;Z_2|Q,Z_1),\right.  \label{firstmax} \\ 
& \left.  \quad I(V';Y_{21}|Q) - I(V';Z_1|Q) - I(V;Z_2|Q, Z_1) \right\} \nonumber
\end{align} 
Now, we note that the terms $I(V';Y_{11}|Q) - I(V';Z_1 | Q)$ and $I(V';Y_{21}|Q) - I(V';Z_1|Q)$ depends only on the marginal distribution $p(q,v',x_1)p(y_{21}, y_{11}, z_1|x_1)$. Similarly, define $Q' = (Q, Z_1)$, the terms $I(V;Y_{12}|Q')  - I(V;Z_2|Q')$ and $I(V;Z_2|Q')$ depends only on the marginal distribution $p(q', v, x_2)p(y_{12}, z_{2}|x_2)$. Therefore, we can further upper bound $R_{CK}$ by
\begin{align*}
R_{\rm CK} \le \max \min & \left\{I(V_1;Y_{11}|Q_1) - I(V_1;Z_1 | Q_1) + I(V_2;Y_{12}|Q_2)  - I(V_2;Z_2|Q_2),\right. \\ 
& \left.  \quad I(V_1;Y_{21}|Q_1) - I(V_1;Z_1|Q_1) - I(V_2;Z_2|Q_2) \right\},
\end{align*}
where the maximum is over $p(q_1)p(v_1|q_1)p(x_1|v_1)$ and $p(q_2)p(v_2|q_2)p(x_2|v_2)$ \footnote{To see that this bound is larger than the previous bound in (\ref{firstmax}), set $V_1 = V'$, $Q_1 =Q$, $V_2 = (V, Q')$ and $Q_2 = Q'$ in this bound to recover the previous bound}. We now further simplify this bound as follows.
\begin{align*}
R_{\rm CK} &\le \max \min \left\{I(V_1;Y_{11}|Q_1) - I(V_1;Z_1 | Q_1) + I(V_2;Y_{12}|Q_2)  - I(V_2;Z_2|Q_2),\right. \\ 
&  \left.  \qquad\qquad \qquad I(V_1;Y_{21}|Q_1) - I(V_1;Z_1|Q_1) - I(V_2;Z_2|Q_2) \right\}, \\
& \le \max \min  \left\{I(X_1; Y_{11}) - I(X_1;Z_1) + I(V_2;Y_{12}|Q_2)  - I(V_2;Z_2|Q_2),\right. \\ 
&  \left.  \qquad \qquad \qquad  I(X_1;Y_{21}) - I(X_1;Z_1) - I(V_2;Z_2|Q_2) \right\},
\end{align*}
where the maximum is now over distributions of the form $p(x_1)$ and $p(q_2) p(v_2|q_2)p(x_2|v_2)$. The last step follows from the fact that $Z_1$ is degraded with respect to both $Y_{21}$ and $Y_{11}$. 
\end{proof}

Next, we evaluate this upper bound. We will make use of the entropy relationship~\cite{Cover}: $H(ap, 1-p,(1-a)p) = H(p,1-p)+pH(a,1-a)$. First consider the terms for the first channel components, \\ $(I(X_1;Y_{11}) - I(X_1;Z_{1}))$ and $(I(X_1;Y_{21}) - I(X_1;Z_{1}))$. Letting $\P\{X_1 = 0\} = \gamma$ and evaluating the individual expressions, we obtain
\begin{align*}
I(X_1;Y_{21}) &= H(\gamma, 1-\gamma), \\
I(X_1;Y_{11}) &= H\left(\frac{\gamma}{2}, \frac{1}{2}, \frac{1-\gamma}{2}\right) - 1 \\
& = \frac{1}{2}H(\gamma, 1-\gamma), \\
I(X_1; Z_{1}) &= H\left(\frac{\gamma}{6}, \frac{5}{6}, \frac{5(1-\gamma)}{6}\right) - H\left(\frac{1}{6}, \frac{5}{6}\right) \\
& = \frac{1}{6}H(\gamma, 1-\gamma).
\end{align*}
This gives
\begin{align*}
I(X_1;Y_{21}) -I(X_1;Z_{1}) &= \frac{5}{6} H(\gamma, 1-\gamma), \\
I(X_1;Y_{11}) -I(X_1;Z_{1}) &= \frac{1}{3} H(\gamma, 1-\gamma).
\end{align*}
Note that both expressions are maximized by setting $\gamma = 1/2$, which yields
\begin{align}
R_{\rm CK} &\leq \min  \left\{\frac{1}{3}+ I(V_2;Y_{12}|Q_2) - I(V_2;Z_{2}|Q_2), \frac{5}{6}- I(V_2;Z_{2}|Q_2)\right\}. \label{eqn:3}
\end{align}

Next, we consider the second channel component terms. Let $\alpha_i = p(q_{2i})$, $\beta_{j,i} = p(v_{2j}|q_{2i})$, $\P\{X_2 = 0|V_2 = v_{2j}\} = \mu_j$, and $\P\{V_{2} = v_{2j}\} = \nu_j$, then
\begin{align*}
I(V_2; Z_{2}|Q_2) & = \sum_i \alpha_i H \left(\frac{\sum_{j}\beta_{j,i}\mu_j}{2}, \frac{1}{2}, \frac{\sum_{j}\beta_{j,i}(1-\mu_j)}{2}\right ) - \sum_{j}\nu_j H\left(\frac{\mu_j}{2}, \frac{1}{2}, \frac{(1-\mu_j)}{2}\right) \\
& = \frac{1}{2}\sum_i \alpha_i H \left(\sum_{j}\beta_{j,i}\mu_j, \sum_{j}\beta_{j,i}(1-\mu_j)\right) - \frac{1}{2}\sum_{j}\nu_j H\left(\mu_j,(1-\mu_j)\right),
\end{align*}
\begin{align*}
I(V_2; Y_{12}|Q_2) & = \sum_i \alpha_i H \left(\sum_{j}\beta_{j,i}\mu_j, \sum_{j}\beta_{j,i}(1-\mu_j)\right) - \sum_{j}\nu_j H\left(\mu_j,(1-\mu_j)\right).
\end{align*}
This implies that
\begin{align*}
I(V_2; Y_{12}|Q_2) - I(V_2; Z_{2}|Q_2) 
& = \frac{1}{2}\sum_i \alpha_i H \left(\sum_{j}\beta_{j,i}\mu_j, \sum_{j}\beta_{j,i}(1-\mu_j)\right) - \frac{1}{2}\sum_{j}\nu_j H\left(\mu_j,(1-\mu_j)\right).
\end{align*}
Comparing the above expressions, we see that $I(V_2;Z_{2}|Q_2) = 0$ implies that $I(V_2; Y_{12}|Q_2) - I(V_2; Z_{2}|Q_2) = 0$. This, together with (\ref{eqn:3}), implies that $R_{\rm CK}$ is \emph{strictly} less than $5/6$.

In comparison, consider the new lower bound in Corollary~\ref{coro1}. Setting $V = X_1$ and $X_1$ and $X_2$ independent Bernoulli $1/2$, we have
\begin{align*}
I(X_1,X_2;Y_{11},Y_{12}) - I(X_1,X_2;Z_{1}, Z_{2}) & = I(X_1;Y_{11}) - I(X_1;Z_{1}) + I(X_2;Y_{12}) - I(X_2;Z_{2}) \\
& = \frac{1}{3} + \frac{1}{2} = \frac{5}{6}, \\
I(V;Y_{2}) - I(V;Z) &= I(X_1; Y_{21}) - I(X_1; Z_{1}, Z_{2}) \\
& = I(X_1; Y_{21}) - I(X_1; Z_{1})  = \frac{5}{6}. 
\end{align*}
Thus, $R = 5/6$ is achievable using the new scheme, which shows that the our lower bound can be strictly larger than the extended Csisz\'{a}r and K\"{o}rner lower bound. In fact, $R = 5/6$ is the capacity for this example since the channel is a special case of the reversely degraded broadcast channel considered in~\cite{Khisti} and we can use the converse result therein to show that $C_{\rm S} \leq 5/6$.

\section{Proof of Theorem  \ref{thm:2}} \label{appen7}
Using Fourier--Motzkin elimination on the rate constraints gives the following region.
 \begin{align}
R_0 &< I(U;Z), \nonumber \\
R_{1}  &< \min\{I(V_0,V_1;Y_1|U) - I(V_1;Z|V_0), I(V_0,V_2;Y_2|U) - I(V_2;Z|V_0)\},\nonumber\\
2R_1 &< I(V_0,V_1;Y_1|U) + I(V_0,V_2;Y_2|U) - I(V_1;V_2|V_0), \\
R_0 + R_1 &< \min\{I(V_0,V_1;Y_1)-I(V_1;Z|V_0), I(V_0,V_2;Y_2) - I(V_2;Z|V_0)\}, \nonumber\\
R_0 + 2R_1  &<  I(V_0,V_1;Y_1)+I(V_0,V_2;Y_2|U)- I(V_1;V_2|V_0), \\
R_0 + 2R_{1} &<  I(V_0,V_2;Y_2) + I(V_0,V_1;Y_1|U)- I(V_1;V_2|V_0),  \\
2R_0 + 2R_1  &<  I(V_0,V_1;Y_1)+ I(V_0,V_2;Y_2)- I(V_1;V_2|V_0),
\end{align}
\begin{align}
R_e &\le R_1, \nonumber \\
R_{e}  &< \min\{I(V_0,V_1;Y_1|U) - I(V_0, V_1;Z|U), I(V_0,V_2;Y_2|U) - I(V_0,V_2;Z|U)\},\nonumber\\
2R_e &< I(V_0,V_1;Y_1|U) + I(V_0,V_2;Y_2|U) - I(V_1;V_2|V_0) - 2I(V_0;Z|U),  \\
R_0 + R_e &< \min\{I(V_0,V_1;Y_1)-I(V_1, V_0;Z|U), I(V_0,V_2;Y_2) - I(V_2, V_0;Z|U)\}, \nonumber\\
R_0 + 2R_e  &<  I(V_0,V_1;Y_1)+I(V_0,V_2;Y_2|U)- I(V_1;V_2|V_0) - 2I(V_0;Z|U), \nonumber\\
R_0 + 2R_{e} &<  I(V_0,V_2;Y_2) + I(V_0,V_1;Y_1|U)- I(V_1;V_2|V_0) - 2I(V_0;Z|U),  \nonumber\\
2R_0 + 2R_e  &<  I(V_0,V_1;Y_1)+ I(V_0,V_2;Y_2)- I(V_1;V_2|V_0) - 2I(V_0;Z|U), 
\end{align}
with the constraint of $I(V_1, V_2;Z|V_0) \le I(V_1;Z|V_0) + I(V_2;Z|V_0) -I(V_1;V_2|V_0)$ on the set of possible probability distributions. Due to this constraint, the numbered inequalities in the above region are redundant. 

We now complete the proof by using rate splitting. This is equivalent to letting $R_1 = R_1''$, $R_0 = R_0^n + R_{1}'$ in the above region and letting the new rates be $R_{0}^n$ for the common message and $R_{1}^n = R_{1}' + R_{11}''$ for the private message. Using Fourier-Motzkin to eliminate the auxiliary rates $R_1'$ and $R_1''$ then results in the following region.

\begin{align*}
R_0 &< I(U;Z), \\
R_0 + R_{1}  &< I(U;Z) +\min\{I(V_0,V_1;Y_1|U) - I(V_1;Z|V_0), I(V_0,V_2;Y_2|U) - I(V_2;Z|V_0)\},\\
R_0 + R_1 &< \min\{I(V_0,V_1;Y_1)-I(V_1;Z|V_0), I(V_0,V_2;Y_2) - I(V_2;Z|V_0)\},
\end{align*}
{\allowdisplaybreaks
\begin{align*}
R_e &\le R_1,  \\
R_{e}  &< \min\{I(V_0,V_1;Y_1|U) - I(V_0, V_1;Z|U), I(V_0,V_2;Y_2|U) - I(V_0,V_2;Z|U)\},\\
R_0 + R_e &< \min\{I(V_0,V_1;Y_1)-I(V_1, V_0;Z|U), I(V_0,V_2;Y_2) - I(V_2, V_0;Z|U)\}, \\
R_0  + 2R_e  &<  I(V_0,V_1;Y_1)+I(V_0,V_2;Y_2|U)- I(V_1;V_2|V_0) - 2I(V_0;Z|U), \\
R_0  + 2R_{e} &<  I(V_0,V_2;Y_2) + I(V_0,V_1;Y_1|U)- I(V_1;V_2|V_0) - 2I(V_0;Z|U), \\
\\
R_0 +R_1+ R_e &< \min\{I(V_0,V_1;Y_1|U) - I(V_1;Z|V_0), I(V_0,V_2;Y_2|U) - I(V_2;Z|V_0)\} \\
& \quad+\min\{I(V_0,V_1;Y_1)-I(V_1, V_0;Z|U), I(V_0,V_2;Y_2) - I(V_2, V_0;Z|U)\}, \\
R_0 +R_1 + 2R_e  &<  \min\{I(V_0,V_1;Y_1|U) - I(V_1;Z|V_0), I(V_0,V_2;Y_2|U) - I(V_2;Z|V_0)\}\\
&\quad + I(V_0,V_1;Y_1)+I(V_0,V_2;Y_2|U)- I(V_1;V_2|V_0) - 2I(V_0;Z|U), \\
R_0 + R_1 + 2R_{e} &<  \min\{I(V_0,V_1;Y_1|U) - I(V_1;Z|V_0), I(V_0,V_2;Y_2|U) - I(V_2;Z|V_0)\}\\
&\quad+ I(V_0,V_2;Y_2) + I(V_0,V_1;Y_1|U)- I(V_1;V_2|V_0) - 2I(V_0;Z|U).
\end{align*}
}
Eliminating redundant inequalities then results in
\begin{align*}
R_0 &< I(U;Z), \\
R_0 + R_{1}  &< I(U;Z) +\min\{I(V_0,V_1;Y_1|U) - I(V_1;Z|V_0), I(V_0,V_2;Y_2|U) - I(V_2;Z|V_0)\},\\
R_0 + R_1 &< \min\{I(V_0,V_1;Y_1)-I(V_1;Z|V_0), I(V_0,V_2;Y_2) - I(V_2;Z|V_0)\},
\end{align*}{\allowdisplaybreaks
\begin{align*}
R_e &\le R_1,  \\
R_{e}  &< \min\{I(V_0,V_1;Y_1|U) - I(V_0, V_1;Z|U), I(V_0,V_2;Y_2|U) - I(V_0,V_2;Z|U)\},\\
R_0 + R_e &< \min\{I(V_0,V_1;Y_1)-I(V_1, V_0;Z|U), I(V_0,V_2;Y_2) - I(V_2, V_0;Z|U)\}, \\
R_0  + 2R_e  &<  I(V_0,V_1;Y_1)+I(V_0,V_2;Y_2|U)- I(V_1;V_2|V_0) - 2I(V_0;Z|U), \\
R_0  + 2R_{e} &<  I(V_0,V_2;Y_2) + I(V_0,V_1;Y_1|U)- I(V_1;V_2|V_0) - 2I(V_0;Z|U), \\
\\
R_0 +R_1 + 2R_e  &<  I(V_0,V_2;Y_2|U) - I(V_2;Z|V_0)+ I(V_0,V_1;Y_1)\\
&\quad +I(V_0,V_2;Y_2|U)- I(V_1;V_2|V_0) - 2I(V_0;Z|U), \\
R_0 + R_1 + 2R_{e} &<  I(V_0,V_1;Y_1|U) - I(V_1;Z|V_0)+ I(V_0,V_2;Y_2)\\
&\quad + I(V_0,V_1;Y_1|U)- I(V_1;V_2|V_0) - 2I(V_0;Z|U).
\end{align*}}

\section{Converse for Proposition~\ref{prop1}} \label{appen4}
The $R_1$ inequalities follow from a technique used in ~\cite[Proposition 11]{Nair}. We provide the proof here for completeness. {\allowdisplaybreaks
\begin{align*}
nR_1 &\leq \sum_{i} I(M_1;Y_{1i}|M_0, Y_{1,i+1}^n) +n\e_n\\
& \leq \sum_{i}I(M_1;Y_{1i}|M_0, Y_{1,i+1}^n, Z^{i-1}) + \sum_i I(Z^{i-1};Y_{1i}|M_0,Y_{1,i+1}^n) +n\e_n \\
& \stackrel{(a)}{\leq} \sum_{i}I(M_1, Y_{1,i+1}^n;Y_{1i}|M_0, Z^{i-1}) -\sum_i I(Y_{1,i+1}^n;Y_{1i}|M_0, Z^{i-1}) \\
& \qquad + \sum_i I(Y_{1,i+1}^n;Z_{i}|M_0,Z^{i-1})+n\e_n \\
& \stackrel{(b)}{\leq} \sum_{i}I(X_i;Y_{1i}|M_0,Z^{i-1}) +n\e_n = \sum_{i}I(X_i;Y_{1i}|U_i) +n\e_n,
\end{align*}
where $(a)$ follows by the Csisz\'{a}r sum lemma; and $(b)$ follows by the assumption that $Y_1$ is less noisy than $Z$ and the data processing inequality. The other inequality involving $Y_2$ and $Z$ can be shown in a similar fashion.}

We now turn to the $R_e$ inequalities. The fact that $R_e \leq R_1$ is trivial. We show the other 2 inequalities. We have
{\allowdisplaybreaks
\begin{align*}
nR_e &\leq I(M_1;Y_1^n|M_0) - I(M_1;Z^n|M_0) +n\e_n\\
& =\sum_{i=1}^n\left( I(M_1;Y_{1i}|M_0, Y_{1,i+1}^n) - I(M_1;Z_{i}|M_0,Z^{i-1})\right) +n\e_n\\
& \stackrel{(a)}{=}\sum_{i=1}^n\left( I(M_1, Z^{i-1};Y_{1i}|M_0, Y_{1,i+1}^n) - I(M_1, Y_{1,i+1}^n;Z_{i}|M_0,Z^{i-1})\right) +n\e_n\\
& \stackrel{(b)}{=}\sum_{i=1}^n\left( I(M_1;Y_{1i}|M_0, Y_{1,i+1}^n, Z^{i-1}) - I(M_1;Z_{i}|M_0,Z^{i-1}, Y_{1,i+1}^n)\right) +n\e_n \\
& \stackrel{(c)}{\le}\sum_{i=1}^n\left( I(M_1, Y_{1,i+1}^n;Y_{1i}|M_0,Z^{i-1}) - I(M_1, Y_{1,i+1}^n;Z_{i}|M_0,Z^{i-1})\right) +n\e_n \\
& \stackrel{(d)}{\le} \sum_{i=1}^n \left(I(X_i;Y_{1i}|U_i)- I(X_i;Z_{i}|U_i)\right) + n\e_n,
\end{align*}
where $(a)$ and $(b)$ follow by the Csisz\'{a}r sum lemma; $(c)$ follows by the less noisy assumption; $(d)$ follows by the less noisy assumption and the fact that conditioned on $(M_0, Z^{i-1})$, $(M_1, Y_{i+1}^n) \to X_i \to (Y_{1i}, Z_i)$. The second inequality involving $I(X;Y_2|U)- I(X;Z|U)$ can be proved in a similar manner. Finally, applying the independent randomization variable $Q \sim \U[1:n],$ i.e. uniformly distributed over $[1:n],$ and defining $U = (U_Q,Q),$ $X = X_Q,$ $Y_{1} = Y_{1Q},$ $Y_{2} = Y_{2Q}$ and $Z = Z_{Q}$ then completes the proof. 

\section{Proof of Proposition~\ref{prop2}}\label{appen6}
In cases two to four, the codebook generation, encoding and decoding procedures are the same as Case 1, but with different rate definitions. We therefore do not repeat these steps here.

\emph{Case 2}: Assume that $I(U_3;Z_3) - R_0 - I(U_3;Z_2|U) \ge 0$, $I(V;Y_1|U_3) - I(V;Z_2|U_3) \le I(V;Y_1|U_3) - I(V;Z_3|U_3)$ and $R_{e3} \le I(V;Y_1|U_3) - I(V;Z_2|U_3)$. \\

In this case, using the definitions of the split message and randomization rates as in case 1, we see that we can achieve $R_{e3} = I(V;Y_1|U_3) - I(V;Z_2|U_3)$ by defining $R_{11}' = I(V;Y_1|U_3) - I(V;Z_2|U_3)$ and $R_{11}'' = 0$. The \textit{equivocation rate constraints} now are
\begin{align*}
R_{10}^o + R_o^r &> I(U_3;Z_2|U), \\
R_1^r + R_{11}^o &> I(V;Z_2|U_3).
\end{align*}
Performing Fourier-Motzkin elimination as before then yields the rate-equivocation region given in Case 2.

\emph{Case 3}: Assume that $I(U_3;Z_3) - R_0 - I(U_3;Z_2|U) \ge 0$, $I(V;Y_1|U_3) - I(V;Z_2|U_3) \ge I(V;Y_1|U_3) - I(V;Z_3|U_3)$ \\
In this case, since we consider only the case of $R_1 \ge I(V;Y_1|U_3) - I(V;Z_3|U_3)$, an equivocation rate of $R_{e3} = I(V;Y_1|U_3) - I(V;Z_3|U_3)$ can be achieved by setting $R_{11}' = I(V;Y_1|U_3) - I(V;Z_3|U_3)$. The constraints for this case are as follow.

\textit{Decoding Constraints}:
\begin{align*}
R_{0} + R_{10}^{o} + R_{0}^r+R_{10}^s &< I(U_3;Z_3), \\
R_{10}^s + R_{10}^o + R_0^r &< I(U_3;Y_1|U), \\
R_{11}' + R_{11}'' + R_{11}^o + R_{1}^r &< I(V;Y_1|U_3).
\end{align*}

\textit{Equivocation rate constraints}:
\begin{align*}
R_{10}^o + R_o^r &> I(U_3;Z_2|U), \\
R_{11}'' + R_1^r + R_{11}^o &> I(V;Z_3|U_3), \\
R_1^r + R_{11}^o &> I(V;Z_2|U_3).
\end{align*}

\textit{Greater than or equal to zero constraints}:
\begin{align*}
R_{10}^o , R_0^o, R_{11}', R_{11}'',R_1^r, R_0^r \ge 0. 
\end{align*}

\textit{Equality constraints}:
\begin{align*}
R_1 &= R_{10}^o + R_{10}^s + R_{11}' + R_{11}'' + R_{11}^o, \\
R_{e2}& = R_{10}^s + R_{11}' + R_{11}'',  \\
R_{e3} &=R_{11}', \\
R_{11}' &=  I(V;Y_1|U_3) - I(V;Z_3|U_3).
\end{align*}

Performing Fourier-Motzkin elimination then results in the rate-equivocation region for Case 3. 

\emph{Case 4}:  Assume that $I(U_3;Z_3) - R_0 - I(U_3;Z_2|U) \le 0$. In this case, note that $R_{e2} \le \min\{R_1, I(V;Y_1|U_3) - I(V;Z_2|U_3)\}$ and can be achieved using only the $V^n$ layer of codewords. We set $R_{10}^s = 0$ in this case. If $ I(V;Y_1|U_3) - I(V;Z_2|U_3) \le  I(V;Y_1|U_3) - I(V;Z_3|U_3)$, then $R_{e2} =  I(V;Y_1|U_3) - I(V;Z_2|U_3)$ and $R_{e3} =\min\{R_1,  I(V;Y_1|U_3) - I(V;Z_3|U_3)\}$ are achievable. If  $I(V;Y_1|U_3) - I(V;Z_2|U_3) \ge  I(V;Y_1|U_3) - I(V;Z_3|U_3)$, then $R_{e3} =  I(V;Y_1|U_3) - I(V;Z_3|U_3)$ and $R_{e2} =\min\{R_1,  I(V;Y_1|U_3) - I(V;Z_2|U_3)\}$ are achievable. 

\section{Proof of Proposition~\ref{prop3}} \label{appen5}
As in~\cite{Nair}, we establish bounds for the channel from $X$ to $(Y_1,Z_2)$ and for the channel from $X$ to $(Y_1,Z_3)$.

\noindent\emph{The $X$  to $(Y_1,Z_2)$ bound}:
{\allowdisplaybreaks
We first prove bounds on $R_0$ and $R_1$. Define the auxiliary random variables $U_{i} = (M_0, Y_1^{i-1})$, $U_{3i} = (M_0, Y_1^{i-1}, Z_{3,i+1}^{n})$, and $V_{i} = (M_1, M_0, Z_{3,i+1}^n, Y_1^{i-1})$ for $i=1,2,\ldots,n$. Then, following the steps of the converse proof in \cite{Abbas3}, it is straightforward to show that
\begin{align*}
R_0 & \leq \frac{1}{n}\sum_{i=1}^{n}I(U_{i};Z_{2i}) + \e_{n}, \\
R_1 & \leq \frac{1}{n}\sum_{i=1}^{n}\left(I(V_{i}; Y_{1i}|U_{i})\right) + \e_{n},
\end{align*}
where $\e_{n} \to 0$ with $n$. 

To bound $R_{e2}$, first consider
\begin{align*}
H(M_1|Z_2^n) & \stackrel{(a)}{\leq} H(M_1|Z_2^n,M_0) + n\e_{n} \\
& \stackrel{(b)}{=} H(M_1) - I(M_1;Z_2^n|M_0) + n\e_{n} \\
& \stackrel{(c)}{\leq} I(M_1;Y_1^n|M_0) - I(M_1;Z_2^n|M_0) + n\e_{n} \\
& = \sum_{i=1}^n(I(M_1;Y_{1i}|M_0, Y_1^{i-1})- I(M_1;Z_{2i}|M_0, Z_2^{i-1})) + n\e_{n} \\
& \stackrel{(d)}{\le} \sum_{i=1}^{n}(I(X_i;Y_{1i}|M_0,Y_1^{i-1}) -I(X_i;Z_{2i}|M_0,Z_2^{i-1})) + n\e_{n} \\
& =  \sum_{i=1}^{n}(I(X_i;Y_{1i}|U_i) -H(Z_{2i}|M_0,Z_2^{i-1})+ H(;Z_{2i}|M_0,Z_2^{i-1}, X_i)) + n\e_{n} \\
& \stackrel{(e)}{\le}  \sum_{i=1}^{n}(I(X_i;Y_{1i}|U_i) -H(Z_{2i}|M_0,Z_2^{i-1}, Y^{i-1})+ H(Z_{2i}|M_0,Z_2^{i-1},Y^{i-1}, X_i)) + n\e_{n} \\
& \stackrel{(f)}{=}  \sum_{i=1}^{n}(I(X_i;Y_{1i}|U_i) -I(X_i;Z_{2i}|U_i)) + n\e_{n},
\end{align*}
where $(a)$ and $(c)$ follow by Fano's inequality, $(b)$ follows by the independence of $M_1$ and $M_0$. $(d)$, $(e)$ and $(f)$ follows by degradation of the channel from $X\to Y_1\to Z_2$, which implies $Z_2^{i-1} \to Y_1^{i-1} \to X_i \to Y_{1i}\to Z_{2i}$ by physical degradedness. For the next inequality, we use the fact that a stochastic encoder $p(x^n|M_0,M_1)$ can be treated as a \emph{deterministic} mapping of $(M_0,M_1)$ and an independent randomization variable $W$ onto $X^n$. 
\begin{align*}
nR_0 + nR_{e2} &= H(M_0) + H(M_1|Z_2^n) \\
&\stackrel{(a)}{\leq} H(M_0) + H(M_1|Z_2^n,M_0) + n\e_n \\
& = I(M_0;Z_3^n) + H(M_1|M_0) -H(M_1|M_0) + H(M_1|Z_2^n,M_0) + n\e_n \\
& = I(M_0;Z_3^n) + I(M_1;Y_1^n|M_0) - I(M_1;Z_2^n|M_0) + n\e_n \\
& \stackrel{(b)}{\leq} I(M_0;Z_3^n) + I(M_1,W;Y_1^n|M_0) - I(M_1,W;Z_2^n|M_0) + n\e_n \\
& \stackrel{(c)}{\leq} \sum_{i=1}^n(I(U_{3i};Z_{3i})+ I(X_i;Y_{1i}|U_{3i}))- I(M_1,W;Z_2^n|M_0) + n\e_n \\
& \stackrel{(d)}{\leq} \sum_{i=1}^n(I(U_{3i};Z_{3i})+ I(X_i;Y_{1i}|U_{3i}))- \sum_{i=1}^n H(Z_{2i}|M_0, Y_{1}^{i-1}) \\
& \qquad\qquad + \sum_{i=1}^n H(Z_{2i}|M_1, M_0, W, Z_{2}^{i-1})+ n\e_n \\
& \stackrel{(e)}{=} \sum_{i=1}^n(I(U_{3i};Z_{3i})+ I(X_i;Y_{1i}|U_{3i}))- \sum_{i=1}^n H(Z_{2i}|M_0, Y_{1}^{i-1})\\
& \qquad\qquad + \sum_{i=1}^n H(Z_{2i}|M_1, M_0, W, Y_{1}^{i-1})+ n\e_n \\
& = \sum_{i=1}^n(I(U_{3i};Z_{3i})+ I(X_i;Y_{1i}|U_{3i}))- \sum_{i=1}^n I(M_1,W, M_0; Z_{2i}|M_0, Y_{1}^{i-1}) + n\e_n \\
& \stackrel{(f)}{=} \sum_{i=1}^n(I(U_{3i};Z_{3i})+ I(X_i;Y_{1i}|U_{3i}))- \sum_{i=1}^n I(X_i; Z_{2i}|M_0, Y_{1}^{i-1}) + n\e_n \\
& = \sum_{i=1}^n(I(U_{3i};Z_{3i})+ I(X_i;Y_{1i}|U_{3i}))- \sum_{i=1}^n I(X_i; Z_{2i}|U_i) + n\e_n, 
\end{align*}
where $(a)$ follows by Fano's inequality and $H(M_0|Z_2^n) \leq n\e_n$; (b) follows by degradation of the channel from $X \to (Y_1,Z_2)$; $(c)$ by Csisz\'{a}r sum applied to the first two terms (see for e.g. \cite{Csiszar}); $(d)$ follows by the fact that conditioning reduces entropy; $(e)$ follows by the Markov relation: $Z_{2}^{i-1} \to Y_1^{i-1} \to (M_0,M_1,W) \to Z_{2i}$; (f) follows by the fact that $X_i$ if a function of $(M_0,M_1,W)$. This chain of inequalities implies that
\begin{align*}
R_{e2} &\leq \frac{1}{n}\sum_{i=1}^n(I(U_{3i};Z_{3i}) - I(U_{3i};Z_{2i}|U_i))-R_0 + \frac{1}{n}\sum_{i=1}^n( I(X_i;Y_{1i}|U_{3i}) -I(X_i; Z_{2i}|U_{3i})) + \e_n \\
& \leq \left[\frac{1}{n}\sum_{i=1}^n(I(U_{3i};Z_{3i}) - I(U_{3i};Z_{2i}|U_i))-R_0\right]^+ + \frac{1}{n}\sum_{i=1}^n (I(X_i;Y_{1i}|U_{3i}) -I(X_i; Z_{2i}|U_{3i})) + \e_n. 
\end{align*}
Finally, we arrive at single letter expressions by introducing the time-sharing random variable $Q \sim \U[1:n],$ i.e. uniformly distributed over $[1:n],$ independent of $(M_0,M_1, X,Y_1,Z_2,Z_3,W)$, and defining $U_{Q} = (M_0, Y_1^{Q-1})$, $U = (U_{Q},Q)$, $V_{Q} = (M_1, U, Z_{2,Q+1}^n)$, $Y_1 = Y_{1Q}$ and $Z_2 = Z_{2Q}$ to obtain the following bounds 
\begin{align*}
R_0 & \leq I(U;Z_{2}) + \e_{n}, \\
R_1 & \leq I(V; Y_{1}|U) + \e_{n},  \\
R_{e2} & \leq (I(X; Y_{1}|U) - I(X; Z_{2}|U)) + \e_{n},\\
R_{e2} &\leq [I(U_3;Z_3)-R_0 - I(U_3;Z_2|U)]^+ + I(X;Y_1|U_3) - I(X;Z_2|U_3) +\e_n.
\end{align*}

\noindent\emph{The $X \to (Y_1,Z_3)$ bound}: The inequalities involving $X \to (Y_1,Z_3)$ follow standard converse techniques. First, applying the proof techniques from \cite{Abbas}, we obtain the following bounds for the rates {\allowdisplaybreaks
\begin{align*}
& R_0 \leq \min \left\{\frac{1}{n}\sum_{i=1}^{n}I(U_{3i};Z_{3i}), \frac{1}{n}\sum_{i=1}^{n}I(U_{3i};Y_{1i})\right\} + \e_{n}, \\
& R_0+R_1  \leq \frac{1}{n} \sum_{i=1}^n (I(V_{i};Y_{1i}|U_{3i}) + I(U_{3i}; Z_{3i})) + \e_{n}.
\end{align*}
We now turn to the second secrecy bound, }
\begin{align*}
H(M_1|Z_3^n) & \leq H(M_1,M_0|Z_3^n) = H(M_1|Z_3^n,M_0) + H(M_0|Z_3^n) \\
& \stackrel{(a)}{\leq} H(M_1|Z_3^n,M_0) + n\e_{n} \\
& \stackrel{(b)}{\leq} H(M_1|Z_3^n,M_0) -H(M_1|Y_1^n,M_0) + n\e_{n} \\
& = I(M_1;Y_1^n|M_0) - I(M_1;Z_3^n|M_0) + n\e_{n}, 
\end{align*}
where $(a)$ and $(b)$ follow by Fano's inequality. Using the Csisz\'{a}r sum lemma, we can obtain the following
{\allowdisplaybreaks
\begin{align*}
H(M_1|Z_3^n) & \leq \sum_{i=1}^n(I(M_1; Y_{1i}|M_0, Y_{1}^{i-1}) - I(M_1; Z_{3i}|M_0, Z_{3, i+1}^{n})) + n\e_n \\
& \stackrel{(a)}{=} \sum_{i=1}^n(I(M_1, Z_{3,i+1}^n; Y_{1i}|M_0, Y_{1}^{i-1})- I(M_1, Y_1^{i-1}; Z_{3i}|M_0, Z_{3, i+1}^{n})) + n\e_n \\
& \stackrel{(b)}{=} \sum_{i=1}^n(I(M_1; Y_{1i}|M_0, Y_{1}^{i-1}, Z_{3,i+1}^n) - I(M_1; Z_{3i}|M_0, Z_{3, i+1}^{n}, Y_1^{i-1})) + n\e_n \\
& = \sum_{i=1}^n (I(V_{i}; Y_{1i}| U_{3i}) - I(V_{i}; Z_{3i}|U_{3i})) + n\e_{n},
\end{align*}
where both $(a)$ and $(b)$ are obtained using the the Csisz\'{a}r sum lemma.
Applying the independent randomization variable $Q \sim \U[1:n]$, i.e. uniformly distributed over $[1:n],$ we obtain
\begin{align*}
R_0 & \leq \min\{I(U_3;Z_3), I(U_3;Y_1)\} + \e_{n}, \\
R_0 + R_1 & \leq I(U_3; Z_3) + I(V; Y_1 | U_3) + \e_{n},\\
R_{e3} & \leq I(V; Y_{1}|U_{3}) - I(V; Z_{3}|U_{3}) + \e_{n},
\end{align*}
where $U_{3Q} = (M_0, Y_1^{Q-1}, Z_{3,Q+1}^n)$, $U_3 = (U_{3Q},Q)$, $Y_1 = Y_{1Q}$ and $Z_3 = Z_{3Q}$. This completes the proof of the outer bound.}
 
\end{document}

%% file: defns.tex
\usepackage{xspace}
\usepackage{mathrsfs}
\usepackage{amsopn}

\newcommand{\Bc}{\mathcal{B}}
\newcommand{\Cc}{\mathcal{C}}
\newcommand{\cc}{\text{\footnotesize $\mathcal{C}$} }

\newcommand{\Ec}{\mathcal{E}}

\newcommand{\Zc}{\mathcal{Z}}

\newcommand{\aep}{{\mathcal{T}_{\epsilon}^{(n)}}}

\newcommand{\lh}{{\hat{l}}}

\newcommand{\Rt}{{\tilde{R}}}

\def\d{\delta}
\def\e{\epsilon}

\let\P\relax
\DeclareMathOperator\P{P}

\newcommand{\U}{\mathcal{U}}

\def\textiid{i.i.d.\@\xspace}
\newcommand\iid{\ifmmode\text{ i.i.d. } \else \textiid \fi}

